\documentclass[10pt]{article} % For LaTeX2e

\usepackage[accepted]{tmlr}
\usepackage{amsmath,amssymb,amsfonts}
\usepackage{cite}
\usepackage{graphicx}
\usepackage{textcomp}
\usepackage{xcolor}
\usepackage{hyperref}
\usepackage{cleveref}
\usepackage{url}
\usepackage{subcaption}
\usepackage{algorithm}
\usepackage{booktabs}
\usepackage{multirow}
\usepackage{multicol}
\usepackage{tabularx}
\usepackage{ragged2e}
\usepackage{algpseudocode}

\usepackage{amsmath,amsfonts,bm}
\usepackage{xspace}

\def\eqref#1{equation~\ref{#1}}
\def\1{\bm{1}}

\DeclareMathAlphabet{\mathsfit}{\encodingdefault}{\sfdefault}{m}{sl}
\SetMathAlphabet{\mathsfit}{bold}{\encodingdefault}{\sfdefault}{bx}{n}

\newcommand{\R}{\mathbb{R}}

\newcommand{\vp}{\bm{p}}               
\newcommand{\vq}{\bm{q}}

\newcommand{\vv}{\bm{v}}       \newcommand{\vvh}{\hat{\bm{v}}}        
               
\newcommand{\vx}{\bm{x}}               
               
\newcommand{\vz}{\bm{z}}       \newcommand{\vzh}{\hat{\bm{z}}}        

\newif\ifLM \LMtrue 
\newif\ifAR \ARtrue \newcommand{\ar}{\ifAR Auto-Regressive (AR)\global\ARfalse\else AR\fi\xspace}
\newif\ifFM \FMtrue \newcommand{\fm}{\ifFM Conditional Flow Matching (FM)\global\FMfalse\else FM\fi\xspace}
\newif\ifVDIFF \VDIFFtrue 
\newif\ifOT \OTtrue 
\newif\ifTTS \TTStrue 
\newif\ifVCF \VCFtrue

\newif\ifWER \WERtrue 
\newif\ifCER \CERtrue 
\newif\ifASR \ASRtrue 
\newif\ifIOU
\IOUtrue
\newcommand{\iou}{\ifIOU Intersection Over Union (IOU)\global\IOUfalse\else IOU\fi\xspace}
\newif\ifFAD \FADtrue \newcommand{\fad}{\ifFAD Fr\'echet Audio Distance~\citep{fad} (FAD)\FADfalse\else FAD\fi\xspace}
\newif\ifKLD \KLDtrue 
\newif\ifSNR \SNRtrue 
\newif\ifBPM
\BPMtrue

\newif\ifPQ \PQtrue \newcommand{\pq}{\ifPQ Production Quality (PQ)\PQfalse\else PQ\fi\xspace}
\newif\ifPRC \PRCtrue \newcommand{\prc}{\ifPRC Production Complexity (PC)\PRCfalse\else PC\fi\xspace}
\newif\ifCE \CEtrue \newcommand{\ce}{\ifCE Content Enjoyment (CE)\CEfalse\else CE\fi\xspace}

\title{Auto-Regressive vs Flow-Matching: a Comparative Study of Modeling Paradigms for Text-to-Music Generation}

\author{\name Or Tal \email or.tal1@mail.huji.ac.il \\
      \addr The Hebrew University \\ Meta Fundamental AI Research \\
      \AND
      \name Felix Kreuk l \email felixkreuk@meta.com \\
      \addr Meta Fundamental AI Research
      \AND
      \name Yossi Adi \email yossi.adi@mail.huji.ac.il\\
      \addr The Hebrew University \\ Meta Fundamental AI Research \\}

\def\month{09}  % Insert correct month for camera-ready version
\def\year{2025} % Insert correct year for camera-ready version
\def\openreview{\url{https://openreview.net/forum?id=xXc5DeaBYw}} % Insert correct link to OpenReview for camera-ready version

\begin{document}

\maketitle

\begin{abstract}
Recent progress in text-to-music generation has enabled models to synthesize high-quality musical segments, full compositions, and even respond to fine-grained control signals, e.g. chord progressions. State-of-the-art (SOTA) systems differ significantly in many dimensions, such as training datasets, modeling paradigms, and architectural choices. This diversity complicates efforts to evaluate models fairly and identify which design choices influence performance the most. While factors like data and architecture are important, in this study we focus exclusively on the modeling paradigm. We conduct a systematic empirical analysis to isolate its effects, offering insights into associated trade-offs and emergent behaviors that can guide future text-to-music generation systems.
Specifically, we compare the two arguably most common modeling paradigms: auto-regressive decoding and conditional flow-matching.
We conduct a controlled comparison by training all models from scratch using identical datasets, training configurations, and similar backbone architectures.
Performance is evaluated across multiple axes, including generation quality, robustness to inference configurations, scalability, adherence to both textual and temporally aligned conditioning, and editing capabilities in the form of audio inpainting. 
This comparative study sheds light on distinct strengths and limitations of each paradigm, providing actionable insights that can inform future architectural and training decisions in the evolving landscape of text-to-music generation.
Audio sampled examples are available at: \url{https://huggingface.co/spaces/ortal1602/ARvsFM}

\end{abstract}

\section{Introduction}
\label{sec:intro}
Unlike text and vision domains, the audio domain, and music generation in particular, has not yet converged on a dominant modeling approach. While both \ar and non-\ar methods have shown strong results, the trade-offs between them remain under explored~\citep{surveymusicai, surveymusicai2, survey_music_multimodal}. In natural language processing, the dominant modeling paradigm is \ar generation over discrete token sequences~\citep{llama,deepseek}.
In computer vision, leading models are typically non-\ar, relying on diffusion or flow-matching processes over continuous latent spaces~\citep{sdxl,instaflow}. However, it is not clear what approach should we follow for music and audio generation. \citet{musicgen, musiclm} demonstrate impressive performance following the \ar approach using discrete audio representation. In contrast, \citet{musicldm, melodyflow} follows the diffusion and flow matching approaches and also show impressive performance. Lastly, \citet{jen1, seedmusic} proposed hybrid methods utilizing \ar and non-\ar methods. 

While a growing number of systems have demonstrated compelling capabilities in text-conditioned music generation, it is unclear what fundamentally accounts for performance differences across models. Variations in training data, latent representations, architecture design, and optimization procedures often confound evaluation. As a result, there is little consensus on whether improvements arise from the modeling paradigm itself or from external factors, like the training data or architectural choices. These contrasts underscore the need for systematic comparison in audio modeling, where foundational choices are still in flux.

To mitigate that, we present a controlled empirical study comparing the two prominent and commonly used approaches for generative modeling in text-to-music generation: \ar and \fm (non-\ar). All models are trained from scratch using the same training data, latent representations, and similar transformer model backbone architectures. We evaluate each modeling paradigm across multiple axes including perceptual quality (\cref{sec:exp1}), adherence to temporal controls (\cref{sec:exp4}), editing capabilities in the form of audio inpainting (\cref{sec:exp5}), inference efficiency (\cref{sec:exp2}) and robustness to training configuration (\cref{sec:exp3}). This design isolates the modeling approach as the primary experimental variable.

\newcommand{\cellwithref}[2]{%
  \begin{minipage}[t]{\linewidth}%
    \justifying #1\vfill\par\smallskip\raggedright\textit{(Sec.~#2)}%
  \end{minipage}%
}
\newcommand{\rowentry}[3]{%
  #1 \textit{(Sec.~\ref{#3})} & #2 \\\midrule
}\newcommand{\entryrow}[3]{%
  #1 \textit{(Sec.~\ref{#3})} & #2 \\
}
\begin{table}[t!]
\caption{A concise summary of our conclusions: Auto-regressive (AR) vs Flow-Matching (FM).}
\label{tab:conclusions}
% \vskip 0.05in
\centering
\renewcommand{\arraystretch}{1.2}
\begin{tabularx}{\textwidth}{>{\justifying\arraybackslash}p{2.2cm}|>{\justifying\arraybackslash}X}
\toprule
\textbf{Axis} & \textbf{Takeaway } \\
\midrule
\rowentry{Text-to-music fidelity}{Both modeling paradigms exhibit comparable performance with a slight favor toward \ar. The chosen latent frame rate shows a large impact over performance regardless of the length of the latent sequence. 
% Last, a tradeoff between the number of inference steps and generation quality exists in the \fm case; requiring a large number of inference steps to maintain comparable performance.
}{sec:exp1}
\rowentry{Control adherence}{\ar follows temporally-aligned conditioning more accurately than \fm though it is prone to accumulated errors (mismatch of melody-chords). Both paradigms demonstrate a controllability–fidelity trade-off.}{sec:exp4}
\rowentry{Music inpainting}{Supervised inpainting \fm yields the smoothest and most coherent edits. A text-to-music \fm model could be used for zero-shot inpainting but would require a hyper-parameter search per-sample or a better sampling strategy to provide more stable outputs.}{sec:exp5}
\rowentry{Inference speed and batch scaling}{Considering inference on a single A$100$ GPU, \ar with KV-cache mainly benefits from scaling the batch size to $\geq64$ for sequence durations $\leq20$ seconds and degrades for longer sequences due to accumulating overheads. This suggest that \ar models would probably be beneficial for systems expecting large demands, e.g. integration of a generative model in social media platforms.
\fm demonstrated faster inference in all other cases for the observed setup.
}{sec:exp2}
\rowentry{Sensitivity to training configuration}{When the number of update steps is capped at $500$k, \fm reaches near-topline (Sec.~\ref{sec:exp1}) quality using batch size $\geq 32$, though its text-match keeps improving with scale. The \ar model needs a larger token budget per update step to match its topline performance. Our observations suggest that both modeling paradigms would benefit from large-scale training, but \fm could offer a more budget-friendly performance trade-off.}{sec:exp3}
\entryrow{Limitations}{This study is centered on a $400$M-parameter transformer and maintained a controlled experimental setup across all evaluations. We acknowledge that alternative sampling strategies, training methods, model architecture or scale could yield different results.}{sec:conclusions}
\bottomrule
\end{tabularx}
\end{table}

Our results highlight consistent differences between the two paradigms, and aims to derive actionable insights to improve future text-to-music generation systems. Auto-regressive models exhibit slightly higher perceptual quality and demonstrate stronger temporally-aligned control adherence, while flow-matching offers faster inference in most cases and also demonstrate better flexibility for editing tasks. Such findings and additional observations outlined in this work provide practical guidance for selecting modeling paradigms in future music generation systems and are broadly covered in the following sections.
Table~\ref{tab:conclusions} draws a summarized overview of our main conclusions.

\section{Related Work}
\label{sec:related}

The idea of generating music through artificial intelligence has evolved significantly, beginning with rule-based symbolic systems~\citep{pinkerton1956information, papadopoulos1999ai, donnelly2011evolving} and progressing to deep-learning approaches capable of synthesizing high-fidelity audio~\citep{musiclm,noise2music,musicgen,jen1,stableaudio}.
Early experiments in AI-driven music creation focused on MIDI-based outputs~\citep{huang2016deep}, with systems like Jukedeck\footnote{\url{https://techcrunch.com/2015/12/07/jukedeck}} offering genre-specific composition based on user-defined prompts.
However, these early models struggled to capture the richness of human-composed music due to their reliance on pre-structured symbolic representations.

The emergence of deep learning revolutionized this field, first introducing deep Recurrent Neural Network systems~\citet{performance-rnn-2017,mao2018deepj}, which further evolved to transformer-based architectures, such as MuseNet~\footnote{\url{https://openai.com/index/musenet/}} and MusicTransformer~\citep{musictransformer}, demonstrating that transformers could generate stylistically coherent compositions. Subsequent works that followed moved beyond MIDI-based approaches to generate raw audio waveforms mainly do so using one of three generative paradigms: \ar decoding, diffusion, and flow-matching. 

A major breakthrough was introduced by JukeBox\citep{jukebox}, which incorporated both instrumental and vocal elements using \ar decoding. 
Following this line of work, \citet{audiolm} introduce AudioLM, which first compresses audio into ``semantic'' and ``acoustic'' tokens and uses an \ar Transformer to predict them. This enables the model to extend a musical excerpt without relying on any symbolic representation. Building on this idea, MusicLM~\citep{musiclm} adds a text-to-semantic stage and a hierarchical \ar decoder, generating $24$kHz music with noticeably better coherence and fidelity than JukeBox. MusicGen~\citep{musicgen} simplifies the pipeline, encoding a $32$kHz audio into EnCodec~\citep{encodec} tokens and trains a single-stage \ar Transformer to predict all token streams jointly, reducing latency while maintaining high prompt adherence and audio quality.

In parallel, inspired by recent success in text-to-image synthesis, diffusion models have emerged as an alternative paradigm, generating music by iteratively refining noise through a learned denoising process. Noise2Music~\citep{noise2music} introduces a cascaded architecture that enables high-fidelity synthesis through progressive upsampling.
MusicLDM~\citep{musicldm}, extends this approach by incorporating beat-synchronous augmentation, improving musical structure alignment with text prompts.
StableAudio~\citep{stableaudio} further refines diffusion-based synthesis by introducing low-latency inference and high-resolution output ($44.1$kHz) offering long generation of full-length songs.

% Flow-Matching Methods:
A more recent development is the adoption of flow-based generative models, which learn a continuous transformation from a simple distribution to the target audio distribution while conditioning on text. AudioBox~\citep{audiobox} introduces a \fm framework capable of handling multiple audio modalities, including music, speech, and environmental sounds.
JASCO~\citep{jasco} refines \fm techniques for music generation, conditioning on textual descriptions and symbolic music features to enhance both coherence and controllability.
MelodyFlow~\citep{melodyflow} optimizes single-stage \fm models for high-fidelity text-guided music generation, improving both efficiency and musical structure adherence.

An additional timely development is the appearance of approaches combining \ar with non-\ar~\citep{lam2023efficient,jen1,seedmusic}, yet this falls outside of the scope of this work and will not be considered. Together, these approaches represent a diverse and rapidly evolving landscape in text-to-music generation. \ar models continue to set strong baselines for musical structure and coherence~\citep{perceiver_ar,music_transformer}, while non-\ar methods, including Diffusion and \fm models, offer promising alternatives for efficient generation and flexible control. Understanding how these paradigms compare under matched conditions remains an open challenge, motivating further exploration.
\section{Background}
\label{sec:bg}

\subsection{Problem Setup and Formulation}
\label{bg:setup}
Given an audio waveform $\vx \in \mathbb{R}^{f_s\cdot t}$ of duration $t$ seconds, sampled at $f_s$ Hz, we assume access to a pre-trained latent representation model that encodes $\vx$ into either (i) a continuous latent representation $\vz \in \mathbb{R}^{D \times f_r \cdot t}$ or (ii) a discrete representation $\vq \in \mathcal{S}^{N_q \times f_r \cdot t}$ (also known as audio tokenization or audio codec). Here, $f_r$ is the latent frame rate, $D$ the latent dimension, $\mathcal{S}$ is the set of discrete code indices and $N_q$ is the number of parallel code streams. Each discrete code stream has its own embedding table, also referred to as codebook. Our EnCodec model encodes $32$kHz music to $50$Hz multi-token stream composed of $4$ codebook streams. The goal is to train a generative model that operates in this latent space, conditioned on a target textual description, using \ar for discrete modeling or \fm for continuous modeling. The generation could also be conditioned on other temporally aligned controls in addition to the textual description, e.g. chord progressions.

\subsection{Auto-Regressive (AR) Decoding}
\label{bg:ar}
The \ar approach models the discrete latent sequence distribution using a causal transformer trained to predict the next token given past context.  
Given an audio segment $\vx$, its textual description $c_{\text{txt}}$, and a discrete latent representation of $\vx$, $\vq \in \mathcal{S}^{N_q \times f_r \cdot t}$, the model is trained to iteratively generate discrete tokens in an auto-regressive manner.
Our discrete latent representation encodes the input waveform $\vx$ to a sequence of $N_q>1$ discrete streams that are obtained by utilizing Residual Vector Quantization (RVQ)~\citep{soundstream,encodec}. 
RVQ quantizes a continuous encoded latent $\vz\in \mathbb{R}^{D \times f_r \cdot t}$ to $N_q$ streams of discrete tokens recursively quantizing the current residual.
Formally, let $\vzh_1$ be the quantized continuous latent representation of $\vz$ for which each temporal entry was replaced with its closest, in terms of euclidean distance, vector in the $1^{\text{st}}$ codebook and let $\vq_1$ be the corresponding indices of that sequence. Then, recursively applying quantization we can define $\forall j\in\{2,...,N_q\}: \vzh_j = \vz - \sum_{l<j}\vzh_l$ and obtain the corresponding indices stream $\vq_j$.
This iterative process then yields the discrete multi-stream representation $\vq=\text{stack}([\vq_1,...,\vq_{N_q}])$.

\paragraph{Training with Delay Pattern.}
Notice, at each timestep, one needs to predict $N_q$ different codes corresponding to different codebooks. This begs the question, \emph{how should we predict these codebooks?}. Naturally, this multi-stream representation dictates an inherent dependence between stream $j$ and all the ones proceeding it. Due to that, predicting $N_q$ codebooks in parallel at each timestep means independently sampling $N_q$ tokens, ignoring the inherently dependent structure. To mitigate that, we follow MusicGen~\citep{musicgen} and train the model using a \textit{delay pattern} to structure the multi-stream discrete representation. Instead of predicting all codebooks simultaneously, each stream is shifted by a predefined delay, enforcing a temporal offset that allows early predictions to condition later ones.
\begin{figure}[t!]
\vspace{-0.5cm}
\centering
\includegraphics[width=0.5\linewidth]{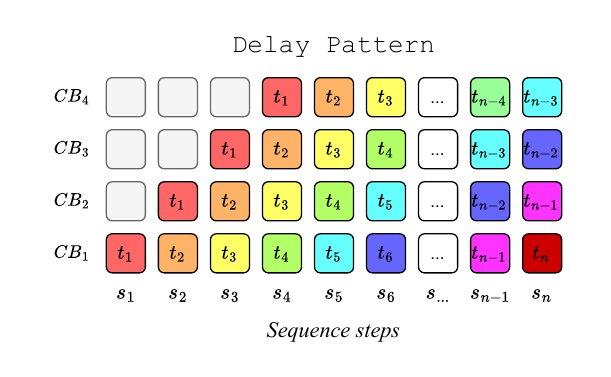}
\vspace{-0.5cm}
\caption{Multi-stream delay pattern modeling. Each row represents a single codebook ($\text{CB}_j$) stream, illustrating the applied delay pattern apperent in the shifting of $\text{CB}_j$ by $j-1$ sequence steps.}
\label{fig:delay pattern}
\end{figure}

For a quantized sequence $\vq \in \mathcal{S}^{N_q \times f_r \cdot t}$ with $N_q$ codebook streams, we impose a delay pattern structure on the multi-stream sequence to allow the $i^{\text{th}}$ codebook at timestep $j$ to be conditioned on all previous $<i$ codebook streams for all $\leq j$ timesteps, as depicted in Figure~\ref{fig:delay pattern}.
Formally, denote $n=f_r \cdot t$, then $\forall i\in[N_q], j\in[n]$, we define a mapping $P: [N_q] \times [n]\rightarrow[N_q] \times [n + N_q - 1]$ s.t $P(i,j)=(i, j + i - 1)$.
The model outputs a distribution over discrete tokens for each codebook stream across the temporal axis $\vp\in\R^{N_q \times |\mathcal{S}| \times f_r \cdot t}$, and trained to minimize the cross-entropy objective:
\begin{equation}
\mathcal{L}_{\text{CE}}(\vq, \vp) = - \frac{1}{N_q \cdot |\mathcal{S}| \cdot f_r \cdot t}\sum_{k\in[N_q]}\sum_{j\in\mathcal{S}}\sum_{i\in[f_r \cdot t]}{\1_\mathrm{\{\vq_{k,i}=j\}}\cdot \log \vp_{k,j,i}}.
\end{equation}

\paragraph{Inference Process.}
At inference time, the model generates tokens sequentially, following the same delay pattern used in training. Generation starts with an empty context, progressively sampling tokens while respecting the predefined temporal shifts between codebooks.
Top-$k$ or top-$p$ sampling strategies are used during decoding, which are standard methods to control diversity in \ar generation. The process continues iteratively until the full latent sequence is generated, after which it is decoded back into a waveform using the pretrained audio tokenizer.

\subsection{Conditional Flow Matching (FM)}
\label{bg:fm}
\fm is an approach to training continuous normalizing flows by regressing a neural network onto a known vector field that generates a probability path~\citep{flow_matching}.
\fm is optimized to predict a deterministic vector field to directly transform noise (or any other input distribution) to the data distribution.

\paragraph{Probability Path Definition.}
We define a continuous transformation from a simple prior $p_0$ (e.g., $\mathcal{N}(0, I)$) to the data distribution $p_1$ using a time-dependent probability flow $\psi_\tau(y | y_1)$.
The probability flow $\psi_\tau(y | y_1)$ defines a continuous path from $y_1$ to $y$ for which $\psi_\tau(y | y_1)=y\sim p_0$ at $\tau=0$ and $\psi_\tau(y | y_1)=y_1\sim p_1$ at $\tau=1$.

Following the optimal transport setup, as defined by~\citet{flow_matching}, we define a probability flow from $p_0$ to $p_1$ such that: (i) the mean and variance evolve linearly with time $\tau \in [0,1]$; (ii) for $y_0 \sim p_0$ and $y_1 \sim p_1$, the probability path $\psi_\tau(y | y_1)$ and its corresponding vector field $\vv_\tau(y | y_1)$ are given by:

\begin{equation}
\begin{aligned}
&\psi_\tau(y|y_1) = (1 - (1-\sigma_\text{min})\tau)y + \tau y_1, \\
% \end{equation}
% \begin{equation}
&\vv_\tau(y|y_1) = \frac{y_1 - (1-\sigma_\text{min})y}{1 - (1-\sigma_\text{min})\tau},
\end{aligned}
\end{equation}
where $\sigma_\text{min}$ is some small constant to ensure numerical stability.

\paragraph{Training Objective.} Rather than modeling the full marginal probability path, the model estimates the conditional vector field $\vvh_\tau(y|y_1)$. Given a text-conditioned latent $y_1$ drawn from $p_1$, we train the model to minimize the mean squared error between the estimated and reference vector fields

\begin{equation}
\mathcal{L}_\text{FM}(\vvh_\tau(y|y_1), \vv_\tau(y|y_1)) = \mathbb{E}_{\tau\sim [0, 1]}\left[ || \vvh_\tau(y|y_1) - \vv_\tau(y|y_1) ||^2\right].
\end{equation}

Following~\citet{jasco} we slightly modify this training objective by applying a $\tau$ dependent loss scaling, assuming a batch size of $B$ samples:
\begin{equation}
\mathcal{L}_\text{FM}(\vvh_\tau(y|y_1), \vv_\tau(y|y_1)) = \frac{1}{B}\sum_{i\in [B]}(1+\tau_i)\cdot|| \vvh_{\tau_i}(y|y_1) - \vv_{\tau_i}(y|y_1) ||^2.
\end{equation}

\paragraph{Inference Process.} 
Generation follows a non-\ar iterative process, iteratively refining the vector field estimation $\vvh(y_\tau|y_1)$ using an ODE solver. Given an initial sample $y_0 \sim p_0$, we update the state iteratively:
\begin{equation}
y_\tau = y_{(\tau - \Delta\tau)} + \Delta\tau \cdot \vvh(y_\tau|y_1).
\end{equation}

While various ODE solvers and sampling schedules could be applied during inference, in this study we only consider two representative sampling methods: Euler’s method for a fixed-grid sampling and Dopri5~\citep{dopri} for a dynamic (adaptive-step) sampling.

\paragraph{Similarity to Diffusion Modeling.} 
Diffusion models~\citep{ddim} and Flow Matching (FM)\citep{flow_matching} are both continuous-time generative modeling frameworks that transform a simple initial distribution (usually Gaussian noise) into a target data distribution.
Despite originating from different theoretical ideas, they end up with similar training procedures and generation processes. 
This similarity means results and “actionable insights” from one paradigm (e.g. improved samplers, better weighting of losses, network architecture tweaks) may directly inform the other~\citep{flow_matching_guide, diffusion_fm_blog}.
For an in-depth derivation of the cases in which this claim applies, refer to Sec. 10 in~\citet{flow_matching_guide}.

\section{Experimental Setup}
\label{sec:exp_setup}
Throughout this paper we consider text-to-music generation tasks. We also evaluate \ar and \fm considering temporally aligned conditioning for music generation and music inpainting. As we perform multiple experiments in this work, this section serves to define the common ground. Deviations from the shared experimental setup defined in this section would be explicitly described in each of the relevant subsections. 

\subsection{Data}
\label{exp_setup:data}
We train our models on a private proprietary dataset containing tracks from Shutterstock~\footnote{\url{https://shutterstock.com/music}} and Pond$5$~\footnote{\url{https://pond5.com}} data collections, which sums to roughly $20$k hours of Mono $32$kHz mixtures paired with textual descriptions.
We evaluate the trained models on a different private proprietary data containing $162$ hours of high-quality Mono $32$kHz mixtures paired with  high quality captioning textual descriptions.
As we wish to observe subtle changes in performance throughout this work, we refrained from evaluating the models over MusicCaps~\citep{musiclm} as it is unclear what subtle differences in performance stand for w.r.t this set due to inconsistencies in audio-quality, sampling rates and audio-text match. For further comparison of the data sources used in this work please see \Cref{apx:data_specs}.

\subsection{Models}
\label{exp_setup:models}
\textbf{Input Representation.}
To isolate the modeling paradigm, we keep the representation space fixed. 
EnCodec's~\citep{encodec} latent provides this bridge as the same encoder outputs: (i) discrete indices for \ar decoding and (ii) continuous vectors for \fm. This design choice follows prior cross-paradigm work, e.g. ~\citet{musicgen, jasco}. 
For completeness, we replicate the entire pipeline using StableAudio's~\citep{stableaudio} open source training code\footnote{\url{https://github.com/Stability-AI/stable-audio-tools/tree/main}} considering matched model sizes and frame-rates thereby checking whether FM benefits from an encoder trained without quantization.
Note, the configurations of both models were adapted to match the desired frame rates and to have comparable model sizes, hence sub-optimal performance of these is possible. The reconstruction quality comparison of the representation models in~\Cref{apx:latent_representation_comp} demonstrates comparable reconstruction quality for both representation models.

% In this study we aim to isolate the modeling paradigm as the observed variable. To do so we use EnCodec~\citep{encodec} to obtain a latent representation of the audio for most of the experiments in this work.
% We chose EnCodec as it was already used in prior works for both \ar~\citep{musicgen} and \fm~\citep{jasco} modeling.
% % EnCodec's~\citep{encodec} quantized discrete representation was used for \ar modeling in and its continuous, pre-quantizer, latent for \fm modeling.
% In addition, to observe how \fm is impacted due to the chosen representation space, we train a VAE-GAN autoencoder following StableAudio's~\citep{stableaudio} open source recipe\footnote{\url{https://github.com/Stability-AI/stable-audio-tools/tree/main}} to compare \fm continuous modeling on both representations.
% Note: the configurations of both models were adapted to match the desired frame rates and to have comparable model sizes, hence sub-optimal performance of these is possible. The reconstruction quality comparison of the representation models in~\Cref{apx:latent_representation_comparison} shows that both representation models demonstrate comparable reconstruction quality.

\textbf{Backbone Model.} For the backbone transformer architecture we use the open source implementation of MusicGen~\citet{musicgen} using a $400$M parameters transformer configuration (specifically - 'musicgen-small'~\footnote{\url{https://github.com/facebookresearch/audiocraft/blob/main/config/model/lm/model_scale/small.yaml}}). For the \fm case we include U-Net-like skip connections, as it is a relatively standard practice in recent years for Diffusion/\fm based approaches that doesn't use the Diffusion transformer (DiT) architecture, e.g. ~\citet{le2023voicebox,jasco,zhang2025diffusion}. 
The inclusion of such skip connections had a notable impact on performance during preliminary experimentation, and therefore we chose to follow this standard practice and add this slight change to the backbone model, only introducing a small number of additional parameters ($\sim7$M). 
We use T$5$~\citep{t5} to obtain text embeddings and pass them via cross-attention layers as text conditions in both cases.
For further details regarding the latent representation models and the backbone transformer architecture refer to \Cref{apx:model specification}.

\subsection{Evaluation Metrics}
\label{exp_setup:metrics}
\paragraph{Perceptual Quality.} We employ \fad with the reference being a high-quality curated proprietary test set. We report \fad using the open source implementation \textit{fadtk}\footnote{\url{https://github.com/microsoft/fadtk}}(vggish)\citep{fadtk_model} where a lower \fad score is associated with a higher perceptual quality.

\paragraph{Audio Aesthetics.} We use Audiobox Aesthetics~\citep{audioboxaesthetics} estimators which serves as a proxy for subjective evaluation, evaluating different properties of the generated audio. Specifically, we consider:
\setlength{\leftmargini}{1cm}
\begin{itemize}
    \item \textit{\pq:} assesses the technical fidelity of the audio, the absence of distortions or other artifacts, well-balanced frequency range and smooth dynamics. It also reflects the skill in recording, mixing, and mastering.
    \item \textit{\prc:} measures the number of audio elements and their interaction. Higher scores indicate layered compositions with multiple sounds, while lower scores reflect simpler, single-source recordings. It also considers how well elements blend together.
    \item \textit{\ce:} captures the subjective appeal of audio, considering emotional impact, artistic skill, and creativity. Higher scores reflect engaging, expressive, and aesthetically pleasing content.
\end{itemize}

\paragraph{Text Description Match.} To evaluate how well the generated audio matches the given textual description we compute the cosine similarity over a joint CLAP~\citep{clap} text-audio representation. The similarity is computed between the track description and the generated audio, measuring audio-text alignment. We use the official pretrained CLAP model (HTSAT-base)\footnote{\url{https://github.com/LAION-AI/CLAP}} in our evaluation.

\paragraph{Temporally-Aligned Control Adherence.} 

We evaluate the adherence to different conditions: (i) chord progressions; (ii) melody; and (iii) drum beat conditioning. We extract pseudo-annotations for these conditions using pretrained classifiers and representation models. Detailed description of how these annotations were obtained can be found in Appendix~\ref{apx:data_preprocess}. We train a model for both \ar and \fm using all conditions together with condition-dropout.

We consider the following evaluation metrics:
\setlength{\leftmargini}{1cm}
\begin{itemize}
    \item \textit{Chords \iou}: We extract chord label and chord switch time in seconds  sequence pairs for both the generated and reference audio waveforms and compute the \iou score between the two. 
    % The considered union corresponds to the unpadded signal frames, and intersection corresponds to the percentage of the unpadded signal that has the same classified <chord label> on both extracted sequences.
    \item \textit{Down-beat F1 score (Beat F1)}: As~\citet{music_controlnet} suggested, we evaluate the down-beat F1 score using \textit{mir eval}\footnote{\url{https://github.com/mir-evaluation/mir_eval}}~\citep{mir_eval} considering a $50$ms tolerance margin around classified downbeats in the reference signal.
    \item \textit{Melody Chromagram Cosine Similarity (Melody Similarity)}: Similarly to~\citet{musicgen} and using the officially released implementation, we convert the resulting audio to a $12$-bins chromagram representation (single octave) and compute the cosine similarity between the reference and the corresponding chromagrams.
\end{itemize}

\subsection{Model Training}
\label{exp1:training}
Unless stated otherwise, we train all models on $10$ second segments for $500$k update steps using a batch size of $256$, AdamW optimizer with a learning rate of $1\cdot10^{-4}$, and a cosine learning rate scheduler with $4000$ steps warmup followed by a cosine learning rate decay. We follow the training objectives as defined in Section~\ref{sec:bg}.

\subsection{Temporally Aligned Conditioning}
\label{bg:cond}
\begin{figure}[t!]
\vspace{-0.5cm}
\centering
\includegraphics[width=0.8\linewidth]{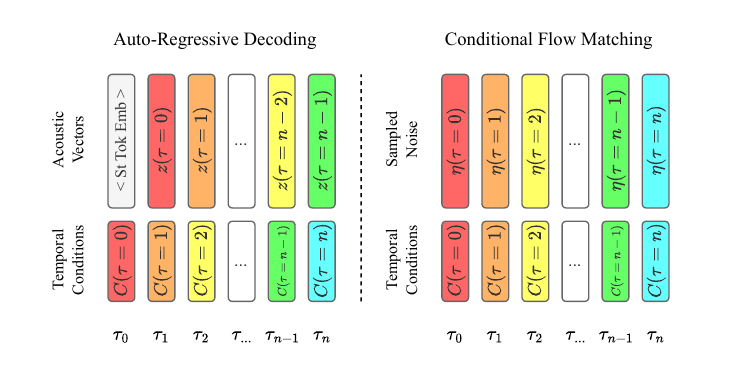}
\vspace{-0.5cm}
\caption{Temporal Conditioning Injection. $\tau_i$ denotes the temporal index in the sequence. In the auto-regressive case we apply a delayed concatenation where the conditions are stacked and concatenated over the channel axis one timestep prior to timestep they correspond to.}
\label{fig:conditioning}
\end{figure}

Following the official release of \textsc{Jasco}~\citep{jasco}, we explore conditioning using temporally aligned controls, i.e. time-dependent conditions. 
Controls are resampled to the latent frame-rate $f_r$ and concatenated to the input signal over the channel axis, then passed through a linear layer prior to the transformer module.

As depicted in \Cref{fig:conditioning} the conditioning injection is done by concatenating the condition vectors (notated as $C$) to the expected transformer input, which is a shifted input in the \ar case (with start token added as the first token) and a sampled standard gaussian noise $\eta$ in the \fm case. The concatenated signals are then projected to the expected transformer dimension using a simple linear projection, and fed as input to the transformer module.

The training objective targets for both modeling paradigms remains the same as the text only variants, i.e. \ar learns next-token prediction, \fm learns the vector field from $\eta$ to~$\vz$. Training is done using a learned new null token for dropout, and inference is done using standard classifier free guidance with unconditional and conditional states.
 
\section{Comparative study: \ar vs \fm}
\subsection{Objective Comparison Under a Fixed Training Setup}
\label{sec:exp1}

We first examine \ar and \fm modeling paradigms under identical conditions.  
Each model trains for one million updates on the same dataset and similar backbone model, and we report objective scores at latent frame rates of $25$, $50$ and $100$ Hz in \Cref{tab:exp1}.
\begin{table}[t!]
\caption{Objective metrics for autoregressive (AR) and flow matching (FM) models after one million updates.  Lower is better for FAD; higher is better for the other metrics.}
\label{tab:exp1}
\centering
\renewcommand{\arraystretch}{1.2}
\begin{tabular}{c|l|ccccc}
\toprule
Hz & Modeling & FAD$\downarrow$ & Clap$\uparrow$ & PQ$\uparrow$& PC$\uparrow$ & CE$\uparrow$ \\ \midrule
% Hz                |  Modeling | FAD  | Clap | PQ  | PC   | CE    | FAD  | Clap | PQ  | PC   | CE    |
\multirow{3}{*}{25} & AR        & 0.40 & 0.41 & 7.71 & 6.02 & 7.36 \\
                    & FM (EnC)  & 0.42 & 0.39 & 7.78 & 5.42 & 7.13 \\
                    & FM (VAE)  & 0.54 & 0.40 & 7.68 & 5.87 & 7.28 \\\midrule
\multirow{3}{*}{50} & AR        & 0.47 & 0.40 & 7.69 & 5.78 & 7.24 \\
                    & FM (EnC)  & 0.48 & 0.40 & 7.73 & 5.60 & 7.20 \\
                    & FM (VAE)  & 0.87 & 0.40 & 7.41 & 5.87 & 7.07 \\ \midrule
\multirow{3}{*}{100}& AR        & 0.64 & 0.40 & 7.59 & 5.84 & 7.17 \\
                    & FM (EnC)  & 0.68 & 0.38 & 7.47 & 5.63 & 6.92 \\
                    & FM (VAE)  & 1.02 & 0.37 & 7.37 & 5.89 & 7.10 \\ \bottomrule
\end{tabular}
\end{table}
\begin{figure}[t!]
\centering
\includegraphics[width=0.9\linewidth]{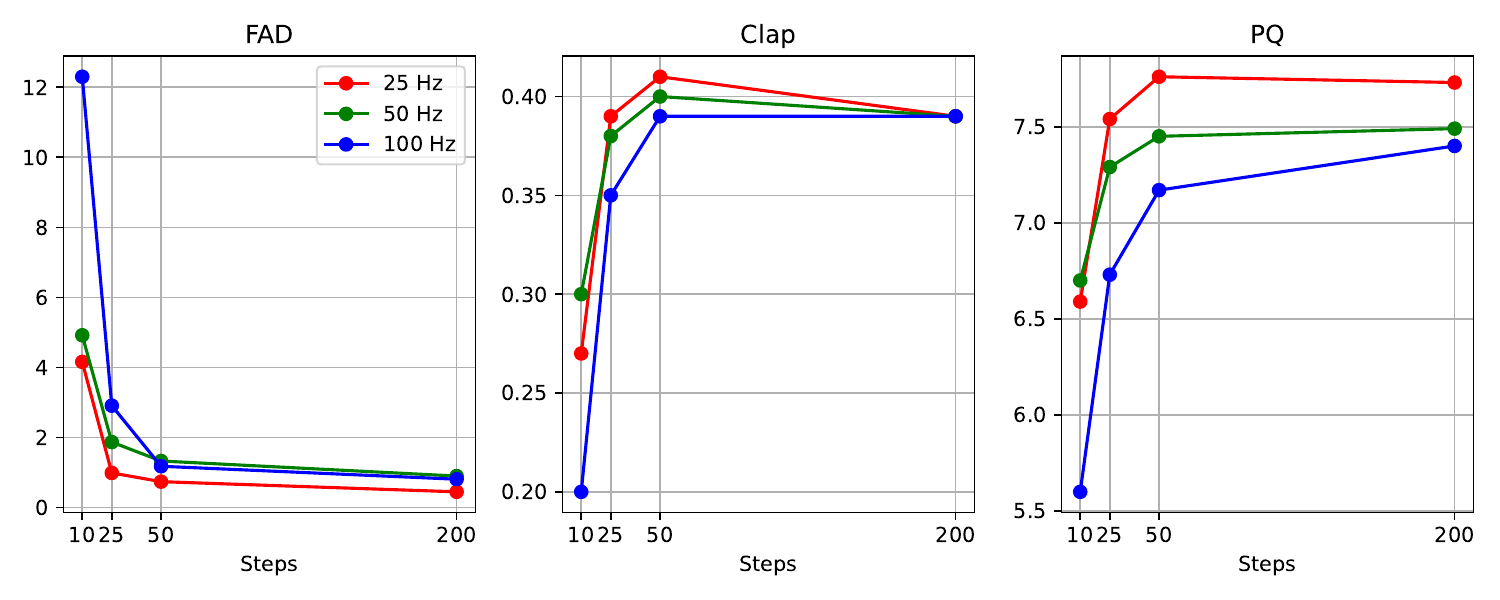}
\caption{\fm performance as a function of inference steps using Euler's method. Decreasing the number of inference steps show a steep degradation in score.}
\label{fig:fm_steps}
\end{figure}
Across the three frame rates, both modeling paradigms exhibit comparable performance, with a slight edge toward \ar in terms of FAD, PC, and CE, while CLAP and PQ scores remain similar between the two.
\ar show less fluctuations as latent frame rate increases, overall and for FAD, PQ and CE in particular,  compared to \fm, especially in the VAE-based latent representation case.
% This observation implies that \ar is more robust to the change in frame-rate.
Interestingly, within the flow matching models, the EnCodec-based model consistently outperform VAE-based one on every metric.
Note, while this observation shows a notable gap in performance, concluding that the EnCodec based space is better for generative modeling is not a direct conclusion.
Within the reasonable exploration of training configurations done in this work, the quantized space (EnCodec based) proved to be better suited for modeling, and it serves to diminish the degrees of freedom in our experiments by using the same representation space for both modeling paradigms.
As mentioned in \Cref{exp_setup:data} - to better observe subtle differences in performance - we perform the observed evaluations on a proprietary high-quality evaluation dataset (see~\Cref{apx:data_specs} for further details). \Cref{apx:exp1_musiccaps} contains the results of this experiment evaluated on the MusicCaps dataset~\citep{musiclm}, demonstrating a noisier and less concrete comparison that mainly serves to set a reference point for the trained model's performance.

\begin{figure}[t!]
\centering
\begin{subfigure}{\textwidth}
    \centering
    \includegraphics[width=0.9\linewidth]{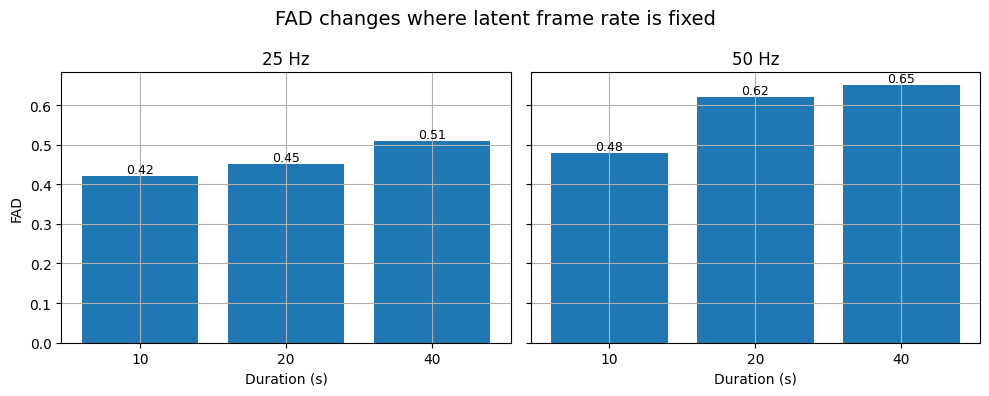}
\end{subfigure}
\begin{subfigure}{\textwidth}
    \centering
    \includegraphics[width=0.9\linewidth]{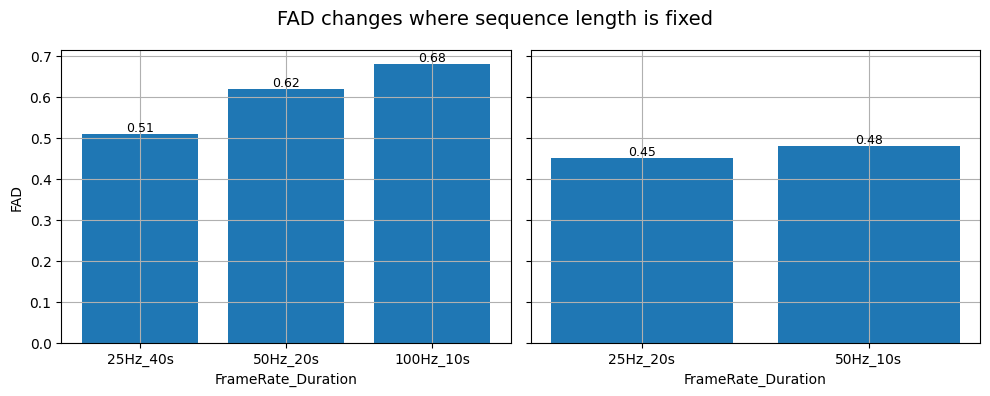}
\end{subfigure}
\caption{Evaluation results of \fm (EnCodec) across varying frame rates (FR) and durations (Dur) - isolating the impact of sequence length and frame rate sensitivity. Full evaluation table is available in \Cref{apx:seq_len_fr}.
Top: fixing latent frame rate and extending the sequence length. Bottom: Fixing sequence length, changing latent frame rate.}
\label{fig:seq_len_fr_abl}
\end{figure}

Moreover, both modeling paradigms demonstrate changes in metrics in correspondence with latent frame rate changes, where \fm shows slightly more apparent fluctuations. 
Such differences could stem from two main factors (i) differences in sequence length (ii) the actual latent representation.
To better understand the nature of this observation we repeat the experiment for $\{25,50\}$Hz \fm (EnCodec) models considering $\{20,40\}$sec durations. 
Under this setup we could isolate the impact of sequence length or the impact of latent frame rate itself; \Cref{fig:seq_len_fr_abl} shows the impact over FAD, and the full results table could be found in~\Cref{apx:seq_len_fr}.
When fixing the latent frame rate, it appears that extending the sequence length has a consistent impact on performance, although the impact is more apparent at $50$Hz than at $25$Hz. 
On the other hand, considering the following pairings: $\{(25\text{Hz}, 20\text{s}), (50\text{Hz}, 10\text{s})\}$, $\{(25\text{Hz}, 40\text{s}), (50\text{Hz}, 20\text{s}), (100\text{Hz}, 10\text{s})\}$ - we fix the length of the sequence the model is required to model; showing a notable gap w.r.t change in representation frame rates.
Both of these observations imply that longer sequences could have an impact on performance, though the representation itself serves as the more crucial factor of the two.
We hypothesize that such implications stem from the nature of the more compressed representation, where less vectors correspond to the same temporal span - resulting in fewer local dependencies.

The best performing \fm configurations follows the Dopri5~\citep{dopri} ODE solver, a dynamic solver that iteratively estimates an error approximation and stops taking additional inference steps if the estimated error is less than a predefined tolerance factor. To further observe how performance is impacted by fixing the number of inference steps we repeat the evaluation using Euler's method on $\{10, 25, 50, 200\}$ inference steps, depicted in \Cref{fig:fm_steps}. As the number of steps reduces below $50$ we see a significant degradation in performance. Taking more steps or using the adaptive Dopri5 solver limits this performance degradation, yet still needs a large number of steps to close the performance gap with \ar. A full evaluation table is available in~\Cref{tab:exp1_fm} on the Appendix.

\textbf{Take-away.} Both modeling paradigms (EnCodec-based latent) show comparable performance with a slight favor toward \ar. The chosen latent frame rate shows a large impact over performance regardless of the length of the latent sequence; an observation that should be taken under consideration when designing a text-to-music generative pipeline. Last, a tradeoff between the number of inference steps and generation quality exists in the \fm case; requiring a large number of inference steps to maintain comparable performance. 
  % fixed training
\subsection{Temporally Aligned Control Adherence}
\label{sec:exp4}
Next, we compare \ar and \fm considering temporally aligned conditioning.
Following JASCO \citep{jasco}, we train $50$ Hz models conditioned on three temporally aligned conditions: chord progression, melody, and drum signals.
As explained in Subsection~\ref{bg:cond} and visualized in Figure~\ref{fig:conditioning}, the injection of the temporally aligned conditions is done by concatenating them over the channel axis prior to the transformer module.
For each paradigm we test two scenarios: (i) all three controls provided; and (ii) one control provided while the others are set to a null token.

\begin{table}[t]
\caption{Adherence to temporally aligned controls.  
“Single” rows report FAD and CLAP for chords/drums/melody conditioning, respectively.}
\label{tab:chords}
\vskip 0.1in
\centering
\renewcommand{\arraystretch}{1.15}
\resizebox{\textwidth}{!}{
\begin{tabular}{l|l|cc|ccc}
\toprule
Modeling & Conditioning & FAD$\downarrow$ & Clap$\uparrow$  & Chords IOU$\uparrow$ & Beat F1$\uparrow$ & Melody Sim.$\uparrow$ \\ \midrule
% Modeling | cnd                     |  FAD  | CLAP | CIOU | Beat | Mld
AR         & \multirow{2}{*}{All}    & \textbf{0.72}  & \textbf{0.37} & \textbf{0.57} & 0.39 & \textbf{0.41} \\
FM         &                         & 0.78  & 0.35 & 0.33 & 0.42 & 0.32 \\ \midrule

AR         & \multirow{2}{*}{Single} & 1.01 / 1.41 / 1.53 & 0.40 / 0.33 / 0.38 & \textbf{0.70} & 0.38 &  \textbf{0.38} \\
FM         &                         & 1.41 / 1.16 / 1.45 & 0.38 / 0.33 / 0.37 & 0.40 & 0.40 &  0.31 \\
\bottomrule
\end{tabular}}
\end{table}

Despite lacking future context, \Cref{tab:chords} shows that the causal \ar decoder tracks the controls more faithfully than \fm. With all streams active, \ar achieves higher Chord IoU (\(0.57\) vs.\ \(0.33\)) and melody similarity (\(0.41\) vs.\ \(0.32\)), while Beat F1 is comparable. The pattern holds in the single-control setting: \ar leads on chords (\(0.70\) vs.\ \(0.40\)) and melody (\(0.38\) vs.\ \(0.31\)), and is on par for drums.  

Interestingly, using temporally aligned conditioning reduces overall fidelity as apparent in \fad and CLAP scores in comparison to the text-to-music model in~\Cref{sec:exp1}.  
Relative to the text-to-music model, \fad rises by \(0.3\)–\(0.8\) and CLAP falls by \(0.02\)–\(0.05\) for both paradigms. We hypothesise that the controls act as a strong bias: once top-\(p\) sampling ventures onto a low-probability path that still satisfies the controls, the model continues down that trajectory, hurting realism. ~\Cref{apx:entropy} expands this observation further.

It appears that the controllability-fidelity tradeoff is enhanced when two of the three controls are dropped, showing a significant increase in chords IOU. Inherently, melody and chords share information (e.g. the musical key) hence such observation is fairly surprising as the controls are given as conditions. One reason the phenomena could stem from is similar to the observation made for quality - a note that doesn't match the chord was generated (audio of that note), be it due to pseudo-labeling error or sampling process, and from that point the local environment is being pushed toward a different chord. The same apply for sampling of a "wrong" chord, or shifts in rhythm. We believe that this observation stems from suboptimal conditioning in this case, where the \ar modeling proves to be more prone to such contradicting controls yet notably adheres better to melody and chords controls.

\textbf{Take-away.} \ar follows temporally-aligned conditioning more accurately than \fm, though it appears to be more prone to accumulated errors (mismatch of melody-chords). Both paradigms lose perceptual quality under strict controls, illustrating a controllability–fidelity trade-off.
  % chord conditioning
\subsection{Inpainting}
\label{sec:exp5}
Music editing often requires replacing a flawed passage while preserving the surrounding context. We therefore compare inpainting capabilities: generating a masked span given past and future audio context. \fm supports na\"{\i}ve zero-shot (ZS) inpainting via latent inversion, whereas \ar does not (at least not for the observed vanilla setup). To enable \ar we adopt the \emph{fill-in-the-middle} strategy of \citet{fill_in_the_middle}, where special tokens split each training example into \(A\!\mid\!B\!\mid\!C\) segments; we present the model with \(A\!\mid\!C\) and ask it to generate \(B\) causally.  For a fair comparison we train both \ar and \fm. We use a fixed $5$ second masked span whose start time is chosen uniformly with at least $1$ second margin on both sides. The algorithms used for \fm inpainting (supervised and ZS) are available in ~\Cref{apx:inpaint}.

\begin{table}[t]
\caption{Objective scores for inpainting with a 5 seconds mask.
Lower is better for FAD; higher is better for all other metrics.}
\label{tab:inpaint}
\centering
\renewcommand{\arraystretch}{1.15}
\begin{tabular}{l|ccccc}
\toprule
Model & FAD$\downarrow$ & CLAP$\uparrow$ & PQ$\uparrow$ & PC$\uparrow$ & CE$\uparrow$ \\
\midrule
AR      & \textbf{0.23} & 0.36          & 7.75          & 5.65          & 7.31 \\
FM      & 0.32          & 0.36          & 7.80          & 5.48          & 7.31 \\
FM (ZS) & 0.30          & \textbf{0.39} & \textbf{7.89} & \textbf{5.73} & \textbf{7.40} \\
\bottomrule
\end{tabular}
\end{table}

\begin{table}[t]
\caption{Human ratings (\(1\!-\!10\); mean \(\pm\) 95\,\% confidence interval, \(\sim\!250\) judgments per cell).}
\label{tab:inpaint_sub}
\centering
\renewcommand{\arraystretch}{1.15}
\begin{tabular}{l|c|ccc}
\toprule
Criterion & GT & AR & FM & FM (ZS) \\
\midrule
Transition smoothness & 8.78$\pm$0.10 & 7.57$\pm$0.19 & \textbf{8.11$\pm$0.15} & 7.09$\pm$0.26 \\
Audio match           & 8.81$\pm$0.12 & 7.22$\pm$0.29 & \textbf{7.93$\pm$0.21} & 6.78$\pm$0.37 \\
\bottomrule
\end{tabular}
\end{table}

\Cref{tab:inpaint} shows that all three approaches achieve similar objective scores.  
However, while listening to the generated audio, we notice artifacts, audible glitches or mismatched timbre, that the objective metrics fail to capture. To support our subjective observations, we conducted a human study, in which the raters were requested to rank each of the observed methods on a scale of $1$ to $10$, where higher is better.
Each audio segment was accompanied with its corresponding waveform figure and a horizontal red line indicator in correspondence to the current temporal position.
The inpainted segment was visibly marked with a \textbf{yellow} background.
For a visualization example see our \href{https://huggingface.co/spaces/Unk-Uname/ARvsFM}{Sample page}. The raters were required to evaluate the following criteria:

\setlength{\leftmargini}{1cm}
\begin{itemize}
    \item \textbf{Transition smoothness}: How smooth the transitions between the yellow (inpainted) and the white (reference) segments is? please refer only to the transitions between the segments.
    \item \textbf{Audio match}: How well does the audio in the yellow segments match the audio in the white segments? A good match should maintain instrumentation, dynamics (volume), tempo, and feel like a part of the same musical piece. Ignore the smoothness of the transition between segments.
\end{itemize}
The results of the human study, presented in \Cref{tab:inpaint_sub}, confirms preliminary observation. Supervised \fm receives the highest scores for both \emph{transition smoothness} and \emph{audio match}, indicating that it generates missing segment with better alignment to the context. \ar ranks second: it produces segments with good fidelity (lowest \fad) but often leaves a discernible seam at the boundaries. Zero-shot \fm delivers the best CLAP, PQ, and CE but exhibits high variance: some samples perfectly fits the context while others drift into unrelated content. This suggests that the sampling configuration could be updated per-sample and a more complex sampling strategy could be used to improve ZS capabilities.

\textbf{Take-away.} Supervised inpainting \fm is the best method among the observed approaches
yielding the smoothest and most coherent edits. Text-to-music \fm could be used
for zero-shot inpainting but would require a hyper-parameter search per-sample or a
better sampling strategy to provide more stable outputs.
  % inpainting
\subsection{Runtime Analysis and Model Scaling}
\label{sec:exp2}

\begin{figure}[t]
\centering
\includegraphics[width=0.90\linewidth]{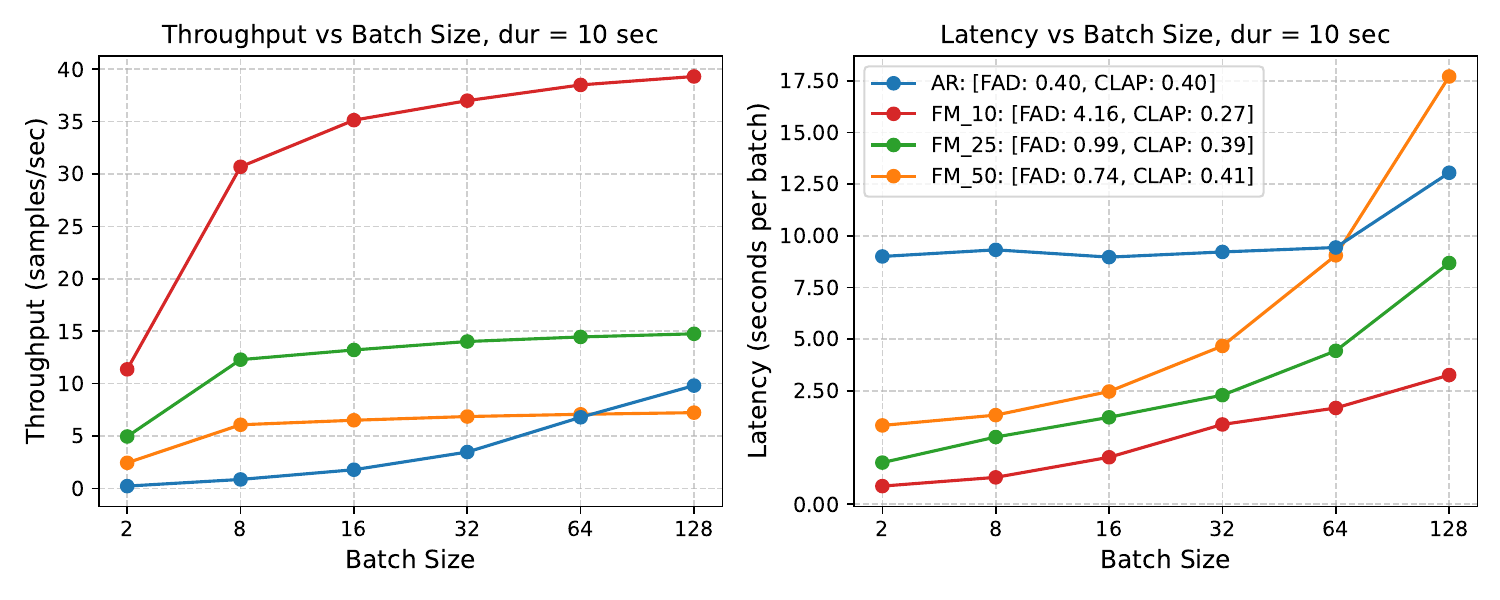}
\caption{Inference speed versus batch size for $10$ sec segments.  
Left: throughput; right: latency. The "Sample" unit refers to a complete generated $10$ sec example.
\ar gains steadily from KV caching, whereas \fm plateaus after batch size $8$.  
Euler's method using $10$ steps is the fastest but has the worst FAD ($4.16$).}
\label{fig:throughput}
\end{figure}

\begin{figure}[t]
\centering
\includegraphics[width=1.0\linewidth]{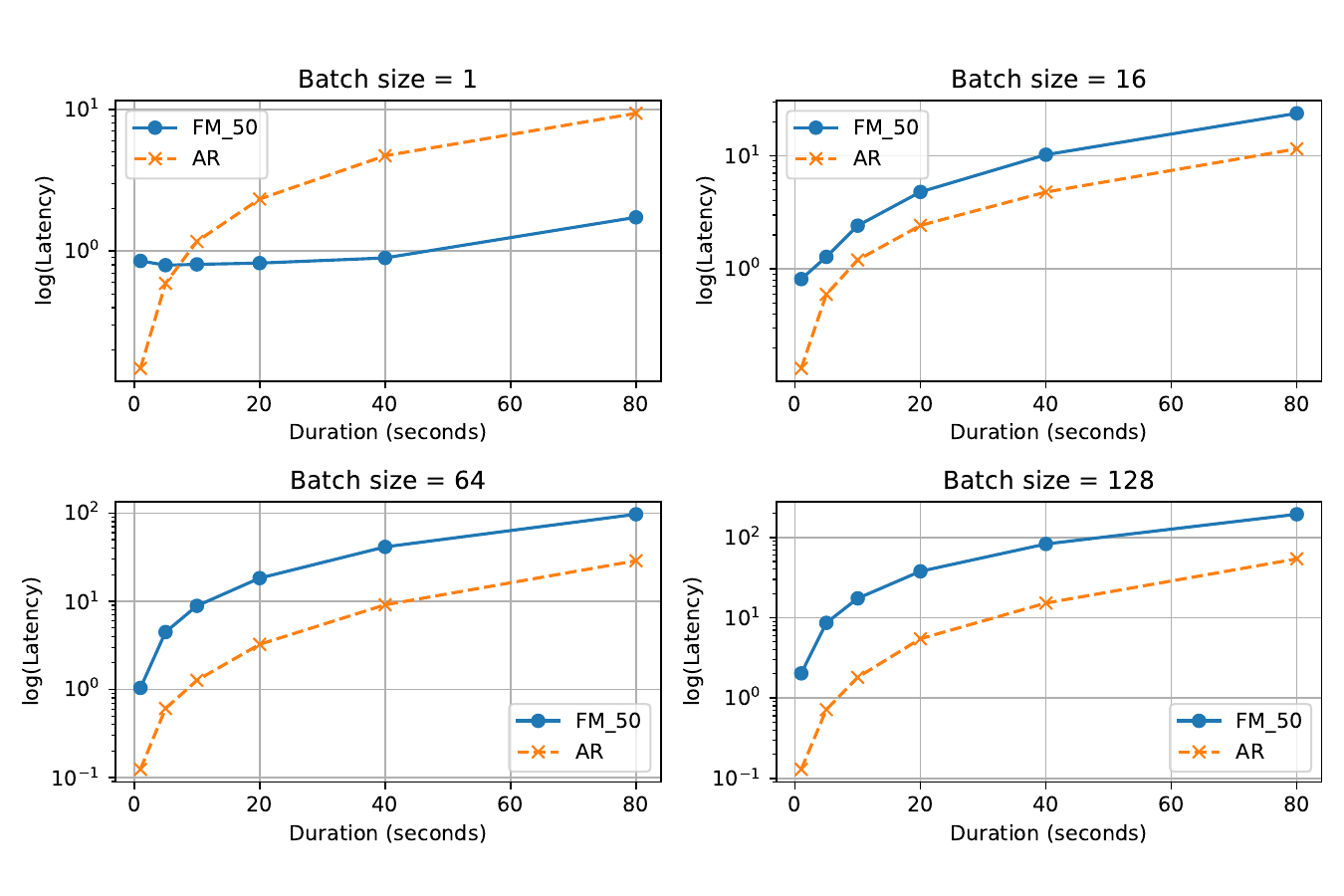}
\caption{Measured latency for fixed batch size and extending sequence duration. Backbone contains $2$ transformer layers.}
\label{fig:latency_vs_dur_2L}
\end{figure}

In subsection~\ref{sec:exp1} we show that \fm can closely match \ar quality when it runs a large number of inference steps and still lags slightly in FAD, PC, and CE.  
This raises two practical questions: (i) \emph{Is it worthwhile to cut the step count to gain inference speed?}; (ii) \emph{Does such speed-up scales to batch processing?}
\ar re-uses hidden states through key–value (KV) caching, so its cost per token falls as the batch grows; \fm has no comparable mechanism. 

To answer these questions, we record throughput (samples / sec) and per-sample latency on a single A$100$ GPU for batch sizes $2$–$256$; in this context "sample" refers to a generated $10$ sec audio sample.
\ar is evaluated with KV caching enabled. \fm is evaluated with Euler's method fixed at $200$, $50$, $25$, and $10$ steps. Objective scores for these settings appear in ~\Cref{apx:alt_fm_steps} (\Cref{tab:exp1_fm}).
\Cref{fig:throughput} shows that the throughput for \ar rises consistently, reaching $6.5$ samples per second at batch $256$. 
In contrast, \fm plateaus: using $50$ steps following Euler's solver, tops out near $3.5$ samples per second. Euler's method using $10$ inference steps is faster than \ar at every batch size, but the audio quality is significantly lower (FAD $4.16$ versus $0.40$), showing a speed-fidelity tradeoff for \fm.

To better understand the behavior during inference, we start by deriving the theoretical run-time complexity expectations. 
Given a batch size $B$, a sequence of length $T$ and latent dimension $D$; we assume that our GPU have some fixed $C$ parallelization capability. The complexity for a full-context attention would then be $O(\frac{BDT^2}{C})$, hence for \fm with $K$ sequential steps we'll get an approximated complexity of $O(K\cdot \max\{\frac{BDT^2}{C}, 1\})$; where $1$ corresponds to the case in which $C$ dominates the fraction. For \ar, using KV caching for self-attention would allow a $O(T)$ cost instead of $O(T^2)$, which yields a single forward pass complexity of $O(\frac{BDT}{C})$. We have to do $T$ sequential steps resulting in a sequence complexity of $O(T\cdot \max\{\frac{BDT}{C}, 1\})$. We then expect to have a $\propto K$ gap in favor of \ar when $C$ does not dominate the fraction.

In practice, this is not the case in the observed setup, as \ar starts to show faster inference than \fm only at batch sizes $>64$. During the derivation above, we assumed that the dominant term to dictate complexity is the self-attention cost, ignoring all other factors. It was claimed in several prior works that \ar inference is often memory-bandwidth bound, not compute-bound~\citep{ar_mem_bound_2,ar_mem_bound_1}, which we hypothesize to be the case in this setup. \Cref{fig:latency_vs_dur_2L} draws latency plots, fixing the batch size and extending the segment duration, considering a small transformer backbone model with 2 transformer layers. The presented latencies for this case correlate with our expectations for run-time complexity. 

\begin{figure}[t]
\centering
\includegraphics[width=1.0\linewidth]{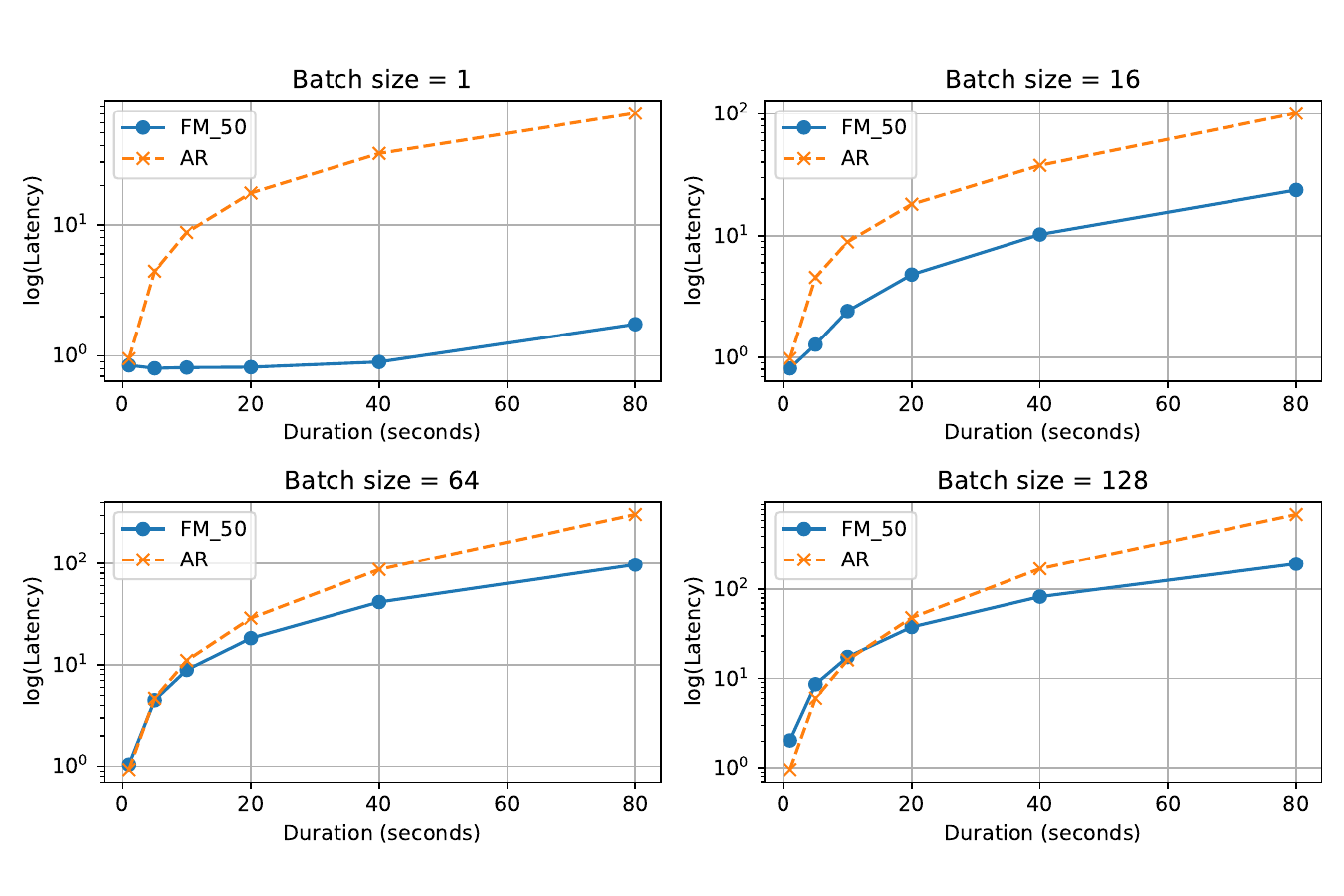}
\caption{Measured latency for fixed batch size and extending sequence duration. Backbone contains $24$ transformer layers.}
\label{fig:latency_vs_dur_24L}
\end{figure}

\Cref{fig:latency_vs_dur_24L} presents the same latency measurement considering $24$ transformer layers for the backbone models. In this case, as we use a single A$100$ GPU, the larger model captures significantly more memory and the measured latencies are significantly degraded for the \ar case; In correlation with \citet{ar_mem_bound_2,ar_mem_bound_1}.
These measurements suggest that in the observed setup, \ar would benefit from KV-caching only for $\leq20$ seconds segments using batch sizes $\geq64$; \fm dominating all smaller batch sizes or longer durations.
To further expand this observation we we a small ablation study, incrementally increasing the number of transformer layers highlighting the accumulating latency as model grows. This experiment could be found in \Cref{apx:runtime}.

\textbf{Take-away.} As seen in this experiment and in \Cref{apx:alt_fm_steps}, there is an apparent tradeoff between model performance and the number of inference steps for \fm in the observed setup.
Considering inference on a single A$100$ GPU, \ar with KV-cache mainly benefits from scaling the batch size to $\geq64$ for sequence durations $\leq20$ seconds and degrades with for longer sequences due to accumulating overheads. This suggest that \ar models would probably be beneficial for systems expecting large demands, e.g. integration of a generative model in social media platforms.
\fm demonstrated faster inference in all other cases for the observed setup.

  % flops, runtime, scaling.
\subsection{Sensitivity to Training Configuration}
\label{sec:exp3}
\begin{figure}[t!]
\centering
\begin{subfigure}{\textwidth}
    \centering
    \includegraphics[width=0.7\linewidth]{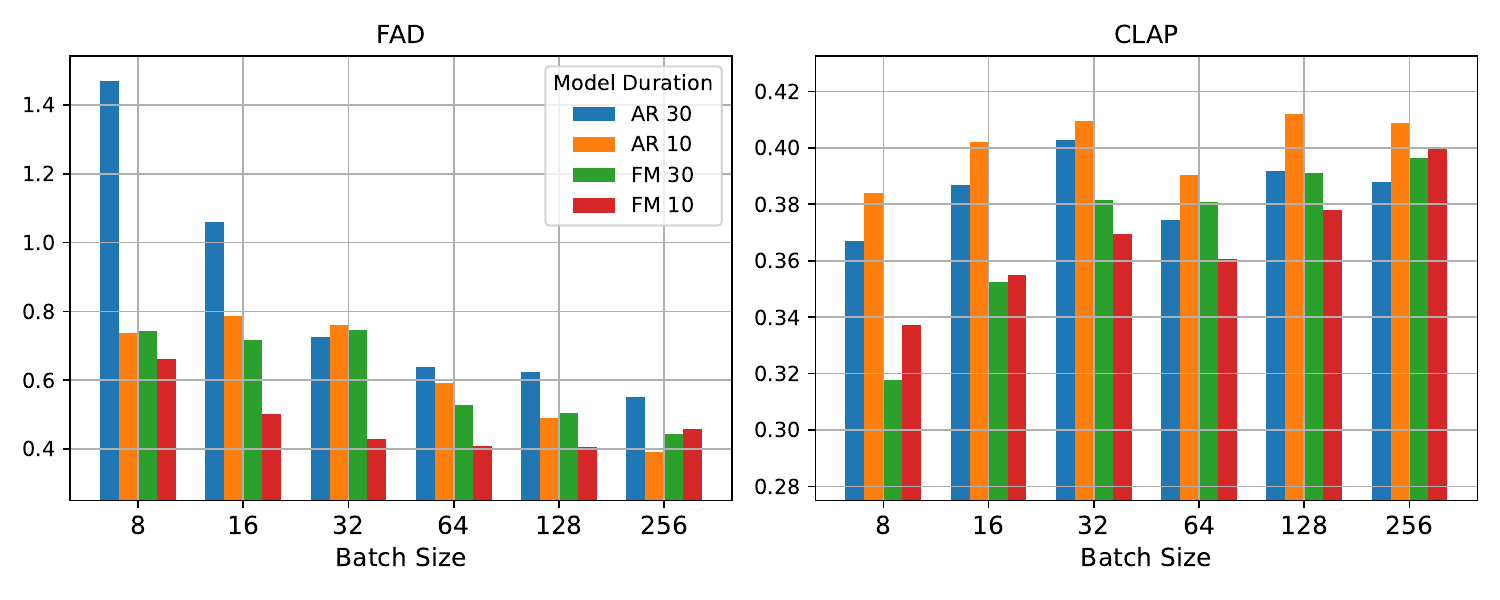}
\end{subfigure}
\begin{subfigure}{\textwidth}
    \centering
    \includegraphics[width=\linewidth]{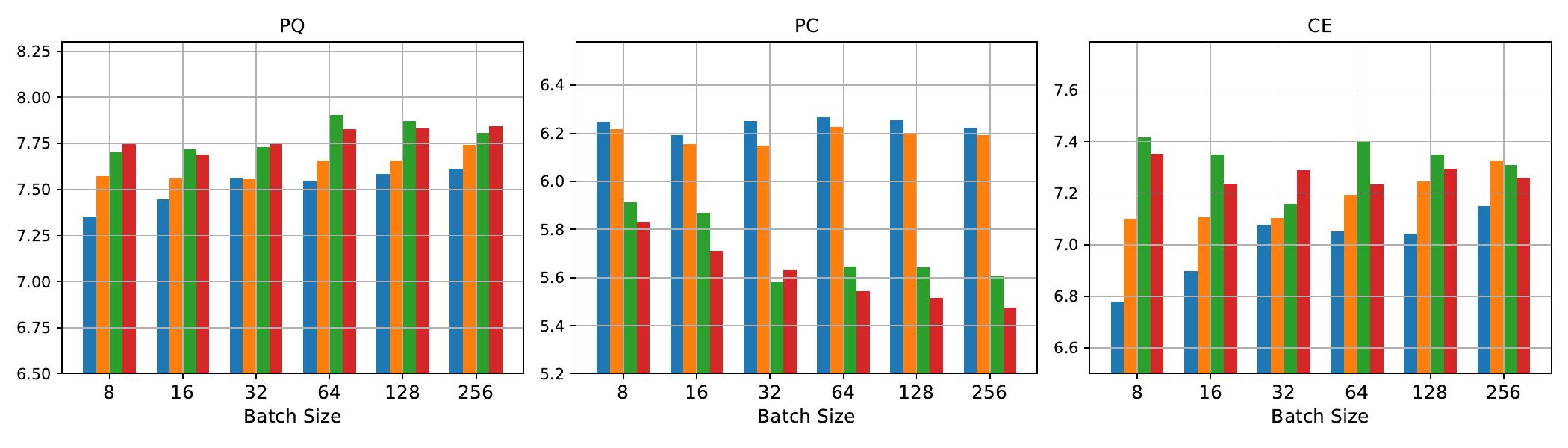}
\end{subfigure}
\caption{Objective scores after $500$k updates as a function of batch size and segment duration.  
Both paradigms improve with more tokens per update step, but \ar is more sensitive to the change.}
\label{fig:exp2_cfg}
\end{figure}
In~\Cref{sec:exp1} we consider models that were trained using a one million update steps and a batch size of $256$. Real-world projects often operate under lower-resource constraints. In this experiment we maintain the number of update steps fixed and vary the \emph{tokens seen per update} by changing batch size and segment duration. The goal is to observe how each modeling paradigm's performance change as a function of batch size and segment duration. For each configuration defined by batch size~$\in\{8, 16, 32, 64, 128, 256\}$ and segment duration~$\in\{10, 30\}[\text{sec}]$ we train two generative models using \ar decoding and \fm.
 
\Cref{fig:exp2_cfg} shows that \fad decreases for both paradigms as batch size increases, where the reduction is steeper for the \ar model. At the largest setting ($256$ batch-size of $10$-second clips) the \ar \fad almost matches the value reached after one million updates, as seen in~\Cref{sec:exp1}, while \fm levels off earlier. When considering CLAP similarity, \fm benefits steadily from larger batches, whereas the \ar shows some fluctuations and relatively flattens above batch size $16$.  Aesthetic metrics present the opposite pattern: PQ and CE are roughly constant for \fm setup, while for the \ar setup they consistently improve. Notice, \ar results are still below the ones obtained after one million update steps, implying there it still benefit from longer training.

\textbf{Take-away.} When the number of update steps is capped, \fm reaches almost the same FAD, PQ, and CE as in the one-million-step topline using batch sizes $\geq 32$, though its CLAP score keeps improving with scale. The \ar model needs a larger token budget per update step to match its topline performance, benefiting more from large scale training. These observations suggest that both modeling paradigms would benefit from large-scale training, but \fm could offer a more budget-friendly performance trade-off.
  % modify training configurations

\section{Conclusion}
\label{sec:conclusions}
This work presents a systematic comparative study of two prominent modeling paradigms for text-to-music generation: Auto-Regressive (AR) decoding and Conditional Flow-Matching (FM). To isolate the effects of modeling choice, we fixed all other factors: training data, latent representation, backbone architecture, and evaluation protocols.
We evaluated both paradigms across five axes: generation quality, control adherence, inpainting capabilities, inference efficiency, and robustness to training configurations.

\paragraph{Key observations}
Our controlled experiments reveal that \ar models demonstrate slightly higher perceptual quality, particularly in Fréchet Audio Distance (FAD), and better adherence to temporally aligned conditioning. \fm models, on the other hand, exhibit strengths in flexibility and editing tasks, notably excelling in supervised inpainting and maintaining inference efficiency in the majority of cases. Notable trade-offs emerged across paradigms: \fm requires a high number of inference steps be comparable to \ar quality, and both paradigms showed a controllability–fidelity trade-off under temporally-aligned conditioning.
For a summary of our conclusions please refer to~\Cref{tab:conclusions},
% We compared \ar and \fm under a unified framework that fixed data, representation model, backbone architecture and evaluation code.
% Five tasks were considered: text-to-music generation at different frame rates, conditioning temporally-aligned controls, inpainting, throughput and latency inference scalability, and sensitivity to training configuration considering a fixed number of training steps.

% \ar achieved the lowest FAD and slightly higher PQ and CE, showing higher robustness to latent frame rate changes; whereas \fm degraded at the higher rate, especially considering a VAE-based latent.
% \ar also followed temporally-aligned controls more closely. General fidelity under this conditional setup decreased for both modeling paradigms, confirming a tension between controllability and generation quality. For inpainting, the supervised \fm model received the highest human ratings for transition smoothness and audio match, while \ar retained the best FAD but often suffered from audible seams; zero-shot \fm showed inconsistent behavior - from seamless natural transitions to generating unrelated segments. On the computational side \ar throughput rose almost linearly with batch thanks to key–value caching, while \fm flattened quickly and surpassed \ar only when considered few steps with an apparent reduction in quality. With a fixed update budget \fm reached near-topline quality with smaller batch sizes, whereas \ar required larger batches and continues to improve from $500$K to $1$M steps.
\paragraph{Limitations and future-work}
This study is centered on a single $400$M-parameter transformer model and maintained a controlled experimental setup across all evaluations. While our findings reveal clear distinctions between \ar and \fm under these constraints, we acknowledge that alternative sampling strategies, efficient training methods, architectural innovations, and model scaling could yield different results. Future work would explore these axes to more comprehensively assess the strengths and limitations of each paradigm. We encourage the community to further investigate \ar and \fm models under fair, unified settings to advance our collective understanding of controllable and efficient music generation.

% \subsection{Broader impact}
\paragraph{Broader impact}
Large scale generative models raise several ethical concerns. To address these, we ensured that all training data was obtained through legal agreements with the appropriate rights holders, primarily through a licensing agreement with the data providers. Generative models may also create challenges for artists by introducing unfair competition, which remains an open issue. We believe that open research is important for providing equal access to all participants, thus we see this research as an important milestone for sharing relevant knowledge not only for researchers in the field but also for anyone working on generative music application. By sharing insights on controlability, e.g. melody conditioning, we believe models developed considering such controls could be valuable tools for both amateur and professional musicians.

\bibliography{main}
\bibliographystyle{tmlr}

\appendix
\section{Dataset specifications}
\label{apx:data_specs}

To date, publicly available human-annotated text-to-music data is limited in terms of quantity, audio-quality, genre diversity and paired textual description quality. To the best of our knowledge - there is not a single publicly available dataset of large quantities of audio files that contains genre-diverse, high quality audio with human annotated text-descriptions. While perhaps there are existing datasets with sufficient quality and quantity (e.g.~\citet{mtgjammendo,jamendomaxcaps}) - obtaining text descriptions for the data would be done by pseudo labeling - an additional free-variable we wished to avoid in this study. To avoid that, we have sourced our data from Pond (\url{https://pond5.com/}) and Shutterstock (\url{https://shutterstock.com/music}). Using these platforms to source the data allows us control over genre diversity, having paired human-labeled descriptions and obtaining high quality audio data.
As for evaluation data, the most common dataset used for evaluation is MusicCaps.
This dataset’s samples are fetched via download from YouTube, which may vary in encoding and sampling rates or recording conditions having a high variance and low textual-match descriptions.
Measuring FAD/CLAP/Aesthetics w.r.t to this set only serves to point out large differences - where it is unclear what subtle differences stand for as this whole set is of lower quality on both audio and text match. 
That being said, using a high quality evaluation set serves to point out finer differences with better reliability - highlighting one method over the other.
To demonstrate text-match and audio quality we employ CLAP cosine similarity and Audiobox-aesthetics metrics over the ground-truth audio and text pairs:

\begin{table}[ht]
\centering
\begin{tabular}{l|c|c|c}
\toprule
\textbf{Dataset} & \textbf{CE} & \textbf{PQ} & \textbf{CLAP} \\
\midrule
Our evaluation set & $7.23 \pm 0.57$ & $7.76 \pm 0.41$ & $0.34 \pm 0.11$ \\
pond 5             & $7.05 \pm 0.79$ & $7.79 \pm 0.48$ & $0.34 \pm 0.12$ \\
Shutterstock       & $7.27 \pm 0.50$ & $7.84 \pm 0.35$ & $0.39 \pm 0.09$ \\
MusicCaps          & $6.14 \pm 1.44$ & $6.91 \pm 1.17$ & $0.28 \pm 0.12$ \\
\bottomrule
\end{tabular}
\label{tab:evaluation}
\end{table}

From these metrics it is apparent that the audio quality suffers from a large variance, where lower PQ implies noisier recordings correlating with CE (subjective evaluation proxy). CLAP values also demonstrate a notable gap in text-audio match. These further outline the advantages of using our suggested datasets for training and evaluation as they offer higher reliability and better confidence in nuanced differences between the observed models as a result.

\section{Latent representation reconstruction comparison}
\label{apx:latent_representation_comp}

We used the training recipe from the open-source implementation of StableAudio to train a VAE-gan representation using a comparable model size and the same latent frame rate, training configuration, and data in order to have an apples-to-apples comparison.

\begin{table}[ht]
\centering
\begin{tabular}{l|cc|cc|cc}
\toprule
Frame-rate & \multicolumn{2}{c|}{\textbf{25}} & \multicolumn{2}{c|}{\textbf{50}} & \multicolumn{2}{c}{\textbf{100}} \\
               & VAE & EnCodec & VAE & EnCodec & VAE & EnCodec \\
\midrule
SISNR          & 4.508 & 5.497 & 7.992 & 8.213 & 11.963 & 10.694 \\
ViSQOL         & 3.859 & 3.879 & 4.062 & 4.005 & 4.191  & 4.111 \\
log-MSD        & 0.414 & 0.404 & 0.339 & 0.351 & 0.279  & 0.314 \\
\bottomrule
\end{tabular}
\label{tab:vae_encodec_metrics}
\end{table}

Both models show relatively comparable reconstruction quality (SiSNR being more sensitive to phase differences); EnCodec was sampled without quantization (pre-quantizer latent was passed to the decoder)

\section{Temporal Controls Preprocessing}
\label{apx:data_preprocess}
In this work we consider drum beat, melody, and chord progression conditioning.
We perform an offline preprocessing stage for each observed latent representation frame rate, and save the preprocessed conditioning signals to memory.
\paragraph{Drum Beat}
To obtain the drum beat supervision we follow ~\citet{jasco}, utilizing a pretrained EnCodec~\citep{encodec} model.
For each data sample, we first encode it to it's corresponding pre-quantization continuous representation using a pretrained EnCodec model that operates in the expected latent representation frame rate.
We then perform temporal blurring~\citep{jasco}, averaging every $5$ sequential latent vectors and broadcast them back to their original frame rate.
Finally, we pass the blurred latent vector through the first vector quantization layer of the pretrained EnCodec model, and save the resulted integer sequence to memory.
\paragraph{Melody}
To obtain the melody condition we use the pretrained deep salience multi-F0 detector \footnote{\url{https://github.com/rabitt/ismir2017-deepsalience}}~\citep{deep_salience}. 
The pretrained multi-F0 detector outputs a confidence score over a predetermined range of $53$ notes (G$2$ to B$7$) spanning over $\sim86$Hz confidence vector sequence.
Given the expected latent representation frame rate, we perform a simple linear interpolation to stretch / shrink the confidence vector stream to match it.
We then pass a threshold of $0.5$ confidence score, zeroing out all values below threshold.
Finally, we create an integer sequence for each data sample, replacing each entry with its corresponding argmax or $54$ in case the column contains only zeros.
\paragraph{Chord Progression}
To obtain chord progressions, we use the Chordino~\footnote{\url{https://github.com/ohollo/chord-extractor}} chord extraction model and create a (<chord label>, <switch time in sec>) pairs sequence for each data sample in our dataset.
Chordino has a vocabulary size of 193 different chords, hence we create a chord to index mapping and save it to memory to be further used for tokenization.
Given an expected latent frame rate, we can then convert the extracted chord sequences to integer sequences using the precomputed chord to index mapping, quantizing the <switch time in sec> timestamps to match the expected frame rate, repeating the same chord index until the next switch.

\section{Sampling Hyperparameter Search}
\label{apx:sampling_hparams}
For both modeling paradigms we experiment with classifier free guidance coefficients $\in \{1.0, 2.0, 3.0, 4.0, 5.0, 6.0, 7.0, 8.0, 9.0\}$.

\noindent For \fm we do inference with Dopri (dynammic number of steps) or Euler (considering number of steps: $\in \{10, 25, 50, 150. 250\})$.

\noindent For \ar we explore temperature $\in\{1.2, 1.4, 1.6, 2.0, 2.4, 2.8\}$, top p $\in \{0.6, 0.8\}$ or top k $\in \{250, 500\}$.

\section{Latent representation and transformer model specifications.}
\label{apx:model specification}

\subsection{Latent Representation}
\label{exp_setup:models:latent}
We follow the approach taken in \citet{musicgen,audiobox,jasco} and use EnCodec's~\citep{encodec} quantized discrete representation for \ar modeling and its continuous, pre-quantizer, latent for \fm modeling.
Using un-normalized representations shows a notable performance gap.
In addition, we train a VAE-GAN autoencoder following StableAudio's~\citep{stableaudio} open source recipe\footnote{\url{https://github.com/Stability-AI/stable-audio-tools/tree/main}}, training with a latent KL divergence constraint w.r.t $\mathcal{N}(0, I)$, without any quantization performed during training.
We train $\{25,50,100\}$[Hz] latent frequency variants for each model, training for $400k$ steps using AdamW optimizer with a learning rate of $3\cdot10^{-4}$ considering $\sim 1$[Sec] segments and a batch size of $64$ samples.
The specific configurations used for each model variant can be found in subsection~\ref{apx:latent_representation_config}.
\subsection{Normalizing EnCodec Latent Representation For FM modeling.}
We sample $N=2048$ $10$ second random segments from our train set and encode them to a latent representation matrix $M$ of shape $[N, T, D]$ where $D$ is the latent dimension and $T$ is the corresponding temporal dimension $T=10\cdot f_r$.
We compute a single scalar for the empirical mean and empirical mean std as follows:
\begin{verbatim}
mean = M.mean()
mean_std = M.std(dim=1).mean()
---
def normalize(z: Tensor):
    return (z - mean) / mean_std

def unnormalize(z: Tensor):
    return z * mean_std + mean
\end{verbatim}

\subsection{Latent Representation Models Configurations}
\label{apx:latent_representation_config}

Both of our observed latent representation models are symmetric auto-encoder models.
We use the open-source implementation in \href{https://github.com/facebookresearch/audiocraft}{Audiocraft} to train EnCodec, and the implementation in \href{https://github.com/Stability-AI/stable-audio-tools}{Stable-Audio-Tools} to train the VAE.
The table below specifies the critical hyper-parameters required to train each of the representation model configurations. In the Discrete EnCodec case we use $4$ residual codebooks, each containing $2048$ bins.
\begin{table*}[ht!]
\centering
\resizebox{\textwidth}{!}{ 
\begin{tabular}{c|ccccc}
\toprule
\textbf{Model} & \textbf{Frame Rate} & \textbf{Strides} & \textbf{Channels} & \textbf{Activation} & \textbf{Latent Dimension} \\
\midrule
\multirow{3}{*}{EnCodec} & 25 & {[}8, 8, 5, 4{]} & \multirow{3}{*}{{[}64, 128, 256, 512{]}} & \multirow{3}{*}{GELU} & \multirow{3}{*}{128} \\
 & 50  & {[}8, 5, 4, 4{]} &  &  &  \\
 & 100  & {[}8, 5, 4, 2{]} &  &  &  \\
 \midrule
\multirow{1.5}{*}{VAE} & 25  & {[}2, 4, 4, 5, 8{]} & \multirow{3}{*}{{[}128, 256, 512, 1024, 2048{]}} & \multirow{3}{*}{Snake} & \multirow{3}{*}{64} \\
 \multirow{2}{*}{$\alpha_{\text{KL}}=10^{-3}$}& 50  & {[}2, 4, 4, 4, 5{]} &  &  &  \\
 & 100 & {[}2, 2, 4, 4, 5{]} &  &  & \\
\bottomrule
\end{tabular}
}
\end{table*}

\subsubsection{Backbone Transformer}
\label{exp_setup:models:trnsformer}
For the backbone transformer decoder model we follow the implementation of ~\citet{musicgen} using a $400$M parameters configuration containing $24$ transformer decoder blocks with a hidden dimension of $1024$, $16$ multi-head attention layers and a feed-forward dimension of $4096$.
We use T5~\citep{t5} to obtain text embeddings and pass them via cross-attention layers as text conditions.
For the \fm case we perform slight modifications similarly to ~\citet{jasco} and include U-Net-like skip connections.
With $2N$ being the number of transformer decoder blocks, we add skip-connections, connecting the input of the $i$'th block with the $2N-i$'th block output, for $i \geq N + 1$.
Each skip connection follows a simple concatenation and linear projection: Linear(Concat($x$, skip)).
The skip connections add $\sim7$M parameters to the network and the input projection for the temporal conditioning experiments add $\sim1$M parameters. The \fm transformer model doesn't require the input embedding tables, reducing $\sim8$M parameters.

\section{Fixed training setup evaluation over MusicCaps}
\label{apx:exp1_musiccaps}
Re-evaluating the experiment presented at \Cref{tab:exp1} on MusicCaps~\citep{musiclm} dataset yields the results depicted in Table~\ref{tab:exp1_MCPS}. The results show a similar trend to \Cref{tab:exp1} where \ar performs better than \fm in most cases, where there is less consistency regarding the representation used for \fm or the degradation of performance as frame rate increases.
As shown in \Cref{tab:vae_encodec_metrics}, MusicCaps suffers from high variance in quality and text-audio match in comparison to our chosen evaluation set - hindering the reliability of the observed trends in this experiment. Therefore, one could view this experimentation as a relative reference point to performance demonstrated by other music-generation works. For example, MusicGen's small configuration had reported to have $3.1$ FAD score using a similar evaluation setup.

\begin{table}[ht!]
\caption{Fixed training configuration evaluation over MusicCaps \label{tab:exp1_MCPS}}
\centering
\renewcommand{\arraystretch}{1.2}
\begin{tabular}{c|l|ccccc}
\toprule
FR & Modeling & FAD$\downarrow$ & CLAP$\uparrow$ & PQ$\uparrow$ & PC$\uparrow$ & CE$\uparrow$ \\
\midrule
\multirow{3}{*}{25} 
& AR        & 4.10  & 0.33 & 7.17  & 5.13 & 6.59 \\
& FM        & 5.53  & 0.30 & 7.42  & 4.17 & 6.42 \\
& FM (VAE)  & 4.58  & 0.30 & 6.99  & 4.70 & 6.24 \\
\midrule
\multirow{3}{*}{50} 
& AR        & 3.55  & 0.33 & 7.10  & 4.93 & 6.36 \\
& FM        & 5.65  & 0.29 & 7.39  & 4.26 & 6.28 \\
& FM (VAE)  & 5.42  & 0.29 & 7.10  & 4.60 & 6.10 \\
\midrule
\multirow{3}{*}{100} 
& AR        & 3.52  & 0.33 & 6.63  & 5.27 & 6.39 \\
& FM        & 6.02  & 0.29 & 7.47  & 4.53 & 6.45 \\
& FM (VAE)  & 4.62  & 0.29 & 7.00  & 5.54 & 6.65 \\
\bottomrule
\end{tabular}
\end{table}

\section{Exploring sensitivity to latent frame rate}
\label{apx:seq_len_fr}

The full evaluation scores for the latent frame rate ablation in~\Cref{sec:exp1}.
Observations shows that both sequence length and the latent representation itself affect performance, with the latter being more crucial.
\begin{table}[ht!]
\centering
\caption{Exploring sensitivity to latent frame rate}
\renewcommand{\arraystretch}{1.2}
\begin{tabular}{c|c|cccc}
\toprule
% FR & Dur (s) & FAD$\downarrow$ & PQ$\uparrow$ & CE$\uparrow$ \\
% \midrule
% \multirow{3}{*}{25} 
% & 10 & 0.42 & 7.78 & 7.13 \\
% & 20 & 0.45 & 7.76 & 7.16 \\
% & 40 & 0.51 & 7.74 & 7.11 \\
% \midrule
% \multirow{3}{*}{50} 
% & 10 & 0.48 & 7.73 & 7.20 \\
% & 20 & 0.62 & 7.65 & 7.12 \\
% & 40 & 0.65 & 7.69 & 7.23 \\
% \midrule
% \multirow{1}{*}{100} 
% & 10 & 0.68 & 7.37 & 7.10 \\
% \bottomrule
% \end{tabular}
% \end{table}
FR & Dur (s) & FAD$\downarrow$ & CLAP$\uparrow$ & PQ$\uparrow$ & CE$\uparrow$ \\
\midrule
\multirow{3}{*}{25} 
& 10 & 0.42 & 0.39 & 7.78 & 7.13 \\
& 20 & 0.45 & 0.39 & 7.76 & 7.16 \\
& 40 & 0.51 & 0.38 & 7.74 & 7.11 \\
\midrule
\multirow{3}{*}{50} 
& 10 & 0.48 & 0.40 & 7.73 & 7.20 \\
& 20 & 0.62 & 0.40 & 7.65 & 7.12 \\
& 40 & 0.65 & 0.39 & 7.69 & 7.23 \\
\midrule
\multirow{1}{*}{100} 
& 10 & 0.68 & 0.38 & 7.37 & 7.10 \\
\bottomrule
\end{tabular}
\end{table}

\section{Constricting Sampling Steps}
\label{apx:alt_fm_steps}
Table~\ref{tab:exp1_fm} depicts the full tradeoff w.r.t objective quality metrics when fixing the number of sampling steps made with Euler ODE solver during \fm inference.
\begin{table}[ht!]
\caption{FM (EnCodec) sampling using fixed number of steps with Euler ODE solver.}
\label{tab:exp1_fm}
\centering
\renewcommand{\arraystretch}{1.2}

\begin{tabular}{c|c|cccccc}
\toprule
Hz & Steps & FAD$\downarrow$ & Clap$\uparrow$ & PQ$\uparrow$& PC$\uparrow$ & CE$\uparrow$ \\ \midrule
% Hz                |  Modeling | FAD  | Clap | PQ  | PC   | CE    |
\multirow{4}{*}{25} & 200       & 0.45 & 0.39 & 7.73 & 5.50 & 7.16 \\
                    & 50        & 0.74 & 0.41 & 7.76 & 5.63 & 7.24 \\   & 25        & 0.99 & 0.39 & 7.54 & 5.48 & 7.01 \\ 
                    & 10        & 4.16 & 0.27 & 6.59 & 5.07 & 5.94 \\\midrule
\multirow{3}{*}{50} & 200       & 0.90 & 0.39 & 7.49 & 5.70 & 6.97 \\
                    & 50        & 1.33 & 0.40 & 7.45 & 5.68 & 6.95 \\
                    & 25        & 1.87 & 0.38 & 7.29 & 5.64 & 6.78 \\ 
                    & 10        & 4.92 & 0.30 & 6.70 & 5.46 & 6.10 \\\midrule
\multirow{3}{*}{100}& 200       & 0.81 & 0.39 & 7.40 & 5.64 & 6.86 \\
                    & 50        & 1.18 & 0.39 & 7.17 & 5.74 & 6.72 \\
                    & 25        & 2.91 & 0.35 & 6.73 & 5.7 & 6.31 \\ 
                    & 10        & 12.3 & 0.20 & 5.60 & 5.02 & 4.67 \\
                    \bottomrule
\end{tabular}
\end{table}

\section{Entropy Analysis Under Strict Temporal Conditioning}
\label{apx:entropy}

To shed light on the quality drop observed in Table~\ref{tab:chords}, we measure how strict conditioning affects the uncertainty of the \ar decoder.  
For each test prompt we record the token probability over the first EnCodec codebook stream at every sampling step and compute its entropy.  
The curve is averaged over 100 random prompts; the text-to-music baseline is obtained by using the models trained in Section~\ref{sec:exp1}.

\begin{figure}[ht]
    \centering
    \includegraphics[width=0.75\linewidth]{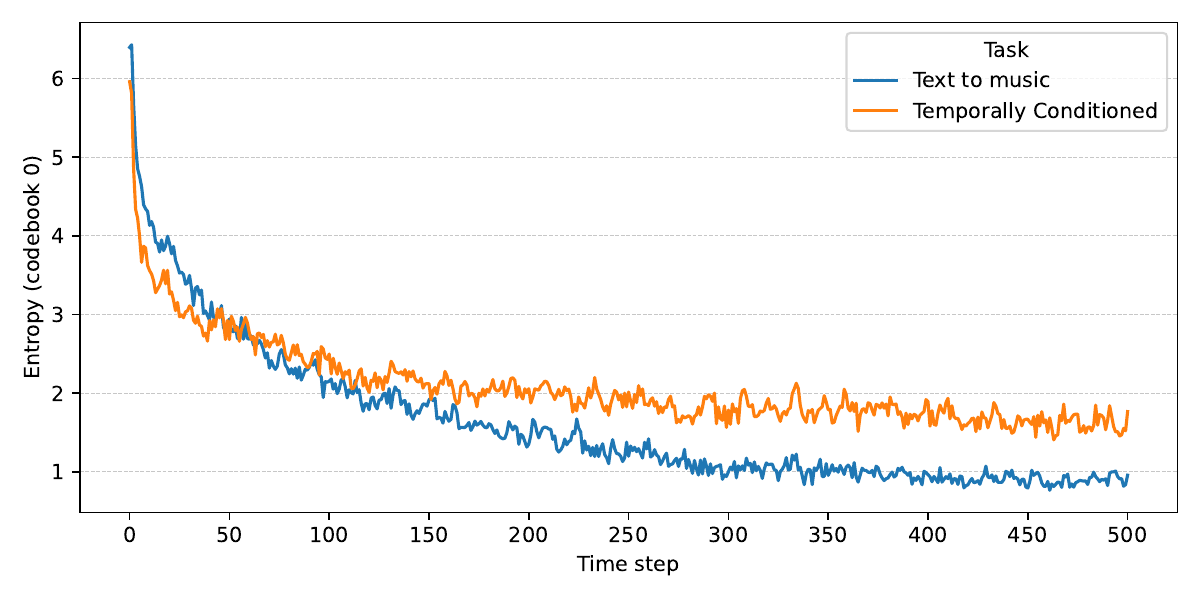}
    \caption{Average entropy of the \ar decoder as a function of sampling step.  
    Conditioning lowers entropy in the first \(\sim\!50\) steps, indicating confident early predictions, but raises it later, suggesting increased sensitivity to low-probability paths.}
    \label{fig:entropy_appendix}
\end{figure}

Figure~\ref{fig:entropy_appendix} shows that conditioning sharpens the distribution early in the sequence, where the provided control tokens directly determine the next few audio tokens, but then the entropy plateaus around timestep $50$.  
We hypothesize that the conditioning imposes a strong bias and therefore leads to a higher chance of sampling out-of-distribution token sequences (some steps doesn't match the conditions), and causes a higher entropy as the sequence becomes longer due to that. 
This manifests as the FAD and CLAP degradation reported in Section-\ref{sec:exp4}.
This also implies that using multi-source classifier-free guidance \citep{jasco} may mitigate this effect, and we leave that for future work.

\section{Inpainting algorithms} 
\label{apx:inpaint}
\begin{algorithm}[ht!]
\caption{Zero-Shot Inpainting Evaluation via Flow Matching Inversion}
\label{alg:zero_shot_inpainting}
\begin{algorithmic}[1]
\Require Source latent $\mathbf{z}_0$ which is a sequence of $T$ latent vectors, condition tensors $\mathcal{C}$, guidance terms $\mathcal{G}$, number of Euler steps $N$ and a fixed mask size $M=T/2$
\Ensure Inpainted latent $\mathbf{z}$
\State \textbf{Initialize:} $\Delta t \gets 1/N$, $\mathbf{z} \gets \mathbf{z}_0$, $t \gets 1$
\State Create empty list \texttt{noises} $\gets [\;]$
\Comment{Perform inversion}
\For{$i = 1$ to $N$}
    \State Compute $\mathbf{v}_\theta \gets$ \texttt{model}$(\mathbf{z}, t, \mathcal{C}, \mathcal{G})$
    \State Update $\mathbf{z} \gets \mathbf{z} - \Delta t \cdot \mathbf{v}_\theta$
    \State Append current $\mathbf{z}$ to \texttt{noises}
    \State Update $t \gets t - \Delta t$
    % \State Compute $\mathbf{v}_\theta \gets$ \texttt{model}$(\mathbf{z}, t, \mathcal{C}, \mathcal{G})$
    % \State Update $\mathbf{z} \gets \mathbf{z} - \Delta t \cdot \mathbf{v}_\theta$
    % \State Append current $\mathbf{z}$ to \texttt{noises}
    % \State Update $t \gets t - \Delta t$
\EndFor
\State Reverse the list \texttt{noises}
\State \textbf{Sample} random start index $s \sim \text{Uniform}(0.1\cdot T, 0.9\cdot T)$
\State Set $e \gets s + M$
\State Initialize $\mathbf{z} \sim \mathcal{N}(0, I)$, $t \gets 0$
\Comment{Perform inpainting}
\For{each noise $\mathbf{n}$ in \texttt{noises}}
    \State Replace $\mathbf{z}[:, :s] \gets \mathbf{n}[:, :s]$, $\mathbf{z}[:, e:] \gets \mathbf{n}[:, e:]$
    \State Compute $\mathbf{v}_\theta \gets$ \texttt{model}$(\mathbf{z}, t, \mathcal{C}, \mathcal{G})$
    \State Update $\mathbf{z} \gets \mathbf{z} + \Delta t \cdot \mathbf{v}_\theta$
    \State Update $t \gets t + \Delta t$
\EndFor
\State Replace $\mathbf{z}[:, :s] \gets \mathbf{z}_0[:, :s]$, $\mathbf{z}[:, e:] \gets \mathbf{z}_0[:, e:]$
\State \Return $\mathbf{z}$
\end{algorithmic}
\end{algorithm}

\begin{algorithm}[ht!]
\caption{Supervised Inpainting Flow Matching}
\label{alg:ft_inpainting_fm}
\begin{algorithmic}[1]
\Require Source latent $\mathbf{z}_0$, condition tensors $\mathcal{C}$, guidance terms $\mathcal{G}$, number of Euler steps $N$ and a fixed mask size $M=T/2$
\Ensure Inpainted latent $\mathbf{z}$
\State \textbf{Sample} random start index $s \sim \text{Uniform}(0.1\cdot T, 0.9\cdot T)$
\State Set $e \gets s + M$
\For{step in $\{1,..., N\}$}
    \State Replace $\mathbf{z}[:, :s] \gets \mathbf{z_0}[:, :s]$, $\mathbf{z}[:, e:] \gets \mathbf{z_0}[:, e:]$
    \State Compute $\mathbf{v}_\theta \gets$ \texttt{model}$(\mathbf{z}, t, \mathcal{C}, \mathcal{G})$
    \State Update $\mathbf{z} \gets \mathbf{z} + \Delta t \cdot \mathbf{v}_\theta$
    \State Update $t \gets t + \Delta t$
\EndFor
\State Replace $\mathbf{z}[:, :s] \gets \mathbf{z}_0[:, :s]$, $\mathbf{z}[:, e:] \gets \mathbf{z}_0[:, e:]$
\State \Return $\mathbf{z}$
\end{algorithmic}
\end{algorithm}

In the fine-tuning case of \ar decoding, we need to introduce $3$ new tokens (<a>, <b>, <c>) in order to partition the source latent representation to $3$ segments: A,B,C.
B is the segment to be inpainted.
We place the special token prior to each segment, resulting in a $T+3$ length segment, and reorganize the sequence as [<a>, A, <c>, C, <b>, B].
During inference we give [<a>, A, <c>, C, <b>] as a prompt to the model, which continues to generate until <eos> or max generation length is met. 
We then re-organize the segments to A, B, C and reconstruct the waveform.

For \fm, we could perform inpainting either by using a pretrained model and perform zero-shot  inpainting, or by training a model specifically for the task.
Algorithm~\ref{alg:zero_shot_inpainting} describes the naive logic implemented to allow for zero-shot inpainting using a \fm model and a fixed sampling schedule.
In the supervised case of \fm, we do not perform inversion, and simply plug in the latent representation itself in the unmasked segments as depicted in algorithm~\ref{alg:ft_inpainting_fm}.

\begin{figure}[ht]
\centering
\includegraphics[width=1.0\linewidth]{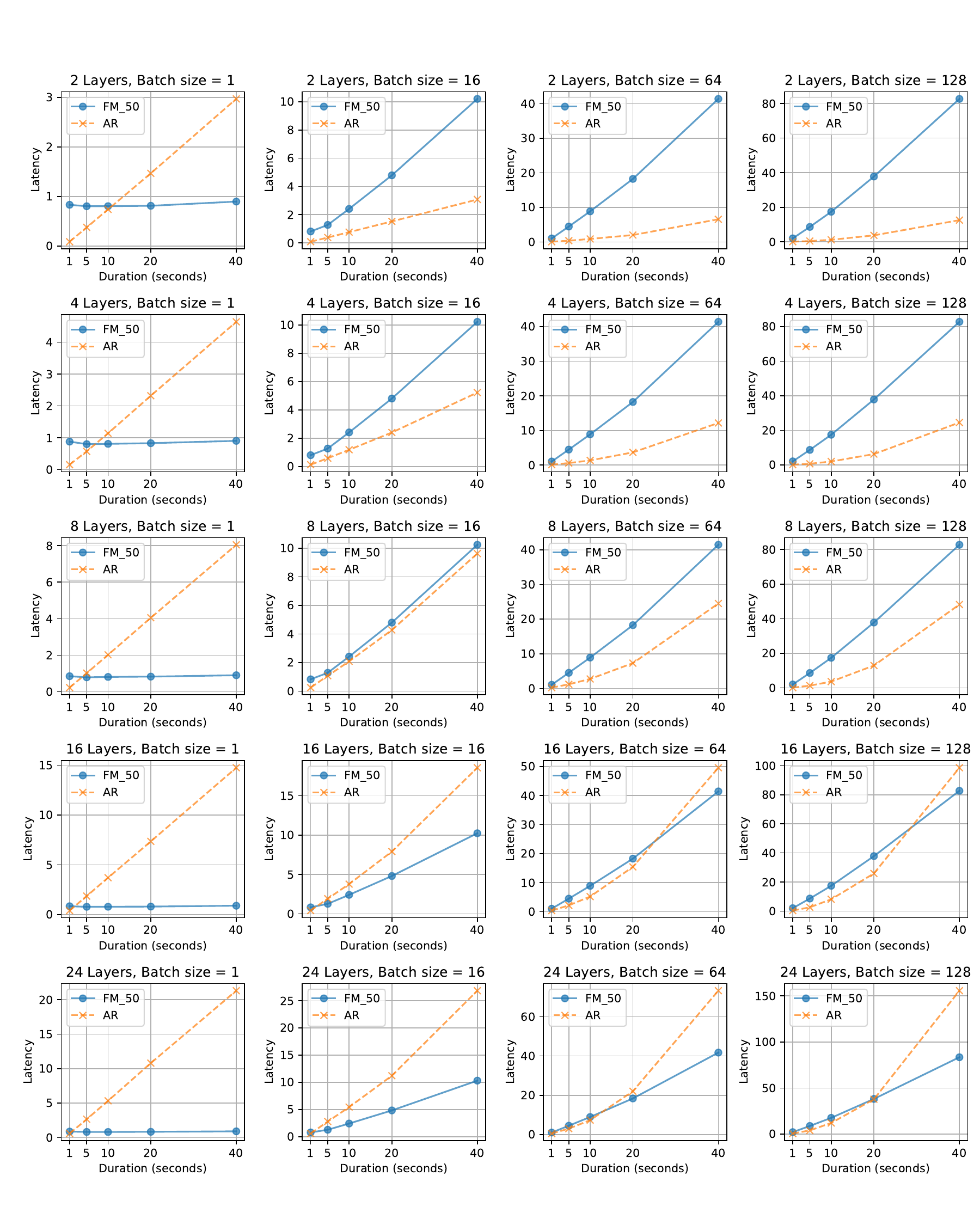}
\caption{Observing the impact of incrementing the number of transformer layers on latency. Column shift from left to right capture batch size increase $2\rightarrow 128$, and row shift from top to bottom represent increase in the number of layers.}
\label{fig:latency_abl}
\end{figure}

\section{Empirical observation over GPU-Memory utilization} 

\label{apx:runtime}
To better capture the the model size - \ar inference speed tradeoff presented in~\Cref{sec:exp2}, we perform a controlled ablation study, considering $[2,4,8,16,24]$ transformer layers in the backbone models in order to observe the tradoff transitioning trends.
To better isolate the runtime overheads, we discard cross attention layers (that can't use KV-cache) and replace the \ar softmax + top-p sampling with argmax sampling.

\Cref{fig:latency_abl} clearly demonstrate the shift in \ar latency, from dominating \fm considering $2$ layers to exponentially exceeding \fm latency.
As expected, longer sequences seem to be increasing in larger latency strides as the performance gap accumulates with $T\gg K$.
Interestingly, despite suffering from large degradations as model size increase - increasing the batch size keeps improving throughput ($\frac{\text{batch size}}{\text{latency}}$) yet with apparent diminishing returns as model size increases, further highlighting the implications of less available memory.

\end{document}